\documentclass[14pt]{extarticle}\usepackage[hyperfootnotes=false]{hyperref}
\usepackage{epsfig}
\usepackage{float}
\usepackage{empheq}
\usepackage{bbold}
\usepackage{cancel}

\usepackage{xspace}

\makeatletter

\@addtoreset{equation}{section}
\makeatother

\usepackage[utf8]{inputenc}
\usepackage{amsmath}
\usepackage{caption}
\usepackage{amsmath}
\usepackage{amssymb}
\usepackage{graphicx}

\newlength{\abstractwidth}
\renewcommand{\thefootnote}{\fnsymbol{footnote}}
\renewcommand{\thanks}[1]{\footnote{#1}} 
\newcommand{\starttext}{
\setcounter{footnote}{0}
\renewcommand{\thefootnote}{\arabic{footnote}}}

\newcommand{\be}{\begin{equation}}
\newcommand{\bea}{\begin{eqnarray}}
\newcommand{\eea}{\end{eqnarray}}
\newcommand{\beq}{\begin{equation}}
\newcommand{\ee}{\end{equation}}

	\newcommand*\widefbox[1]{\fbox{\hspace{2em}#1\hspace{2em}}}

	\def\dsp.{de Sitter space.}

	\def\la{\langle}
	\def\ra{\rangle}
	\def\simleq{\; \raise0.3ex\hbox{$<$\kern-0.75em
			\raise-1.1ex\hbox{$\sim$}}\; }
	\def\simgeq{\; \raise0.3ex\hbox{$>$\kern-0.75em
			\raise-1.1ex\hbox{$\sim$}}\; }
	
	\def\bi{\begin{itemize}}
		\def\ei{\end{itemize}}

	\def\bsub{ \begin{subequations}
			\begin{empheq}[box=\widefbox]{align}  }
			\def\esub{ \end{empheq}
	\end{subequations}}
	
	\def\1{\(  \mathbb{1} \)}

	\def\dk{${\rm DSSYK_{\infty}}$}
	
	\makeatletter
	\g@addto@macro\normalsize{%
		\setlength\abovedisplayskip{10pt}
		\setlength\belowdisplayskip{20pt}
		\setlength\abovedisplayshortskip{10pt}
		\setlength\belowdisplayshortskip{20pt}
	}
	\makeatother
	
	\usepackage{color}

	\usepackage{authblk}
	\title{\Large \bf Holograms and Standard Models}

	\author[1]{Shoichiro Miyashita}
	\author[1]{Yasuhiro Sekino}
	\author[2,3]{Leonard Susskind}
	\affil[1]{Department of Liberal Arts and Sciences,
Faculty of Engineering, Takushoku University, 
Hachioji, Tokyo 193-0985, Japan \vspace{1em}}
	\affil[2]{Stanford Institute for Theoretical Physics and Department of Physics, Stanford University, Stanford, CA 94305-4060, USA \vspace{1em}}
	\affil[3]{Google, Mountain View, CA, USA}
	\date{}
	\usepackage{upgreek}

\begin{document}
		
		\begin{titlepage}
			\maketitle
			
			\begin{abstract}
Despite riddles, mysteries, and enigmas the theory of elementary particles has not changed for 50 years. We argue that the current paradigm -- Wilsonian quantum field theory -- is inadequate for resolving the puzzles, and should be replaced by a new paradigm based on the Holographic Principle. We show how a simplified but still very rich standard model emerges from de Sitter holography in the flat-space limit and explain why the puzzle of huge quantum corrections to the cosmological constant simply does not occur in the holographic paradigm.

\end{abstract}

		\end{titlepage}
		
		\rightline{}
		\bigskip
		\bigskip\bigskip\bigskip\bigskip
		\bigskip
		
		\starttext \baselineskip=17.63pt \setcounter{footnote}{0}

		
		\tableofcontents

\section{Introduction} \label{Introduction}
This paper is about the Holographic Principle; not AdS/CFT but in more cosmologically relevant de Sitter space. However we will not be speaking so much about gravity as about particle physics. By particle physics we mean the degrees of freedom that remain active in the flat-space limit with energies kept fixed and Newton's constant $G$ going to zero. Presumably the real-world Standard Model is such a limit of a holographic theory, but we will use the term standard model in a more general sense. 

\subsection*{A Bold Claim:}
The Holographic Principle gives an entirely new approach to some of the thorniest puzzles of high energy physics -- especially the fine-tuning problems that have plagued the field for more than half a century \cite{Zeldovich:1967gd}. According to the Holographic Principle \cite{tHooft:1993dmi, Susskind:1994vu} a hologram at the boundary of a region of space must do more than just describe gravity in that region. It has to encode everything: a spectrum of fields; particles; masses; coupling constants and forces. In other words a ``standard model'' (SM). 
\begin{figure}[h]
	\begin{center}
		\includegraphics[scale=.4]{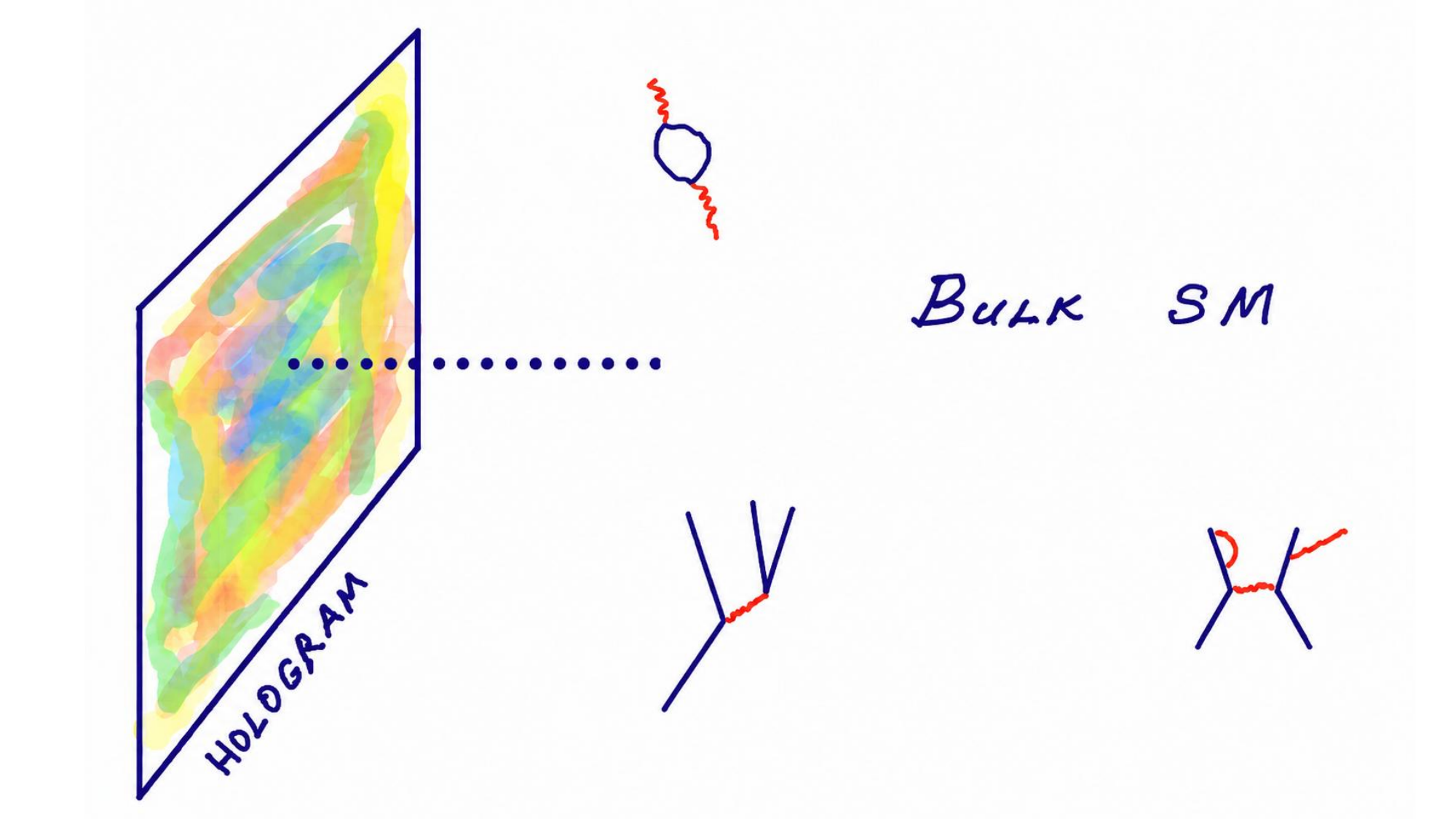}
		\caption{A standard model (SM) and its hologram.}
		\label{Fig1}
	\end{center}
\end{figure} \\
Not necessarily {\textcolor{red}{\underline{\textcolor{black}{the}}}} Standard Model but {\textcolor{red}{\underline{\textcolor{black}{a}}}} standard model -- the bulk physics that emerges from the hologram. Phenomena such as electromagnetism, weak interactions, Higgs, quark confinement, chiral symmetry breaking, CP violation, hadron spectroscopy, even the dimension of space; all of these must be holographically encoded, {\textcolor{red}{\underline{\textcolor{black}{even in the flat-space limit}}}} $G\to0$ ($N\to \infty$). We repeat: even in the flat-space limit $G\to 0$. In other words a SM, if it is the flat-space limit of a de Sitter theory must itself be holographic. This has implications that we will discuss.

It's an aspirational but largely unachieved ambition -- to start with a hologram and to derive its particle physics. In practical terms it seems well beyond the state of the art. Well, maybe -- maybe not. 

We're going to show you a ``toy'' SM that emerges from a toy hologram. The hologram and the SM can be called toys but they are both highly non-trivial. The SM has features closely resembling the real Standard Model: A $U(1) \times SU(N)$ gauge theory, aka QED $\times$ QCD; electric forces; quarks; mesons; baryons; Regge trajectories; confinement. And it resolves the fine-tuning problem -- the instability of hierarchies -- in an unexpected holographic way, {\textcolor{red}{\underline{\textcolor{black}{without supersymmetry}}}}.

Bold Claims! We will justify them, at least for the toy model.

\section{Static Patch Holography}
Let's begin with Static Patch Holography for JT-de Sitter.
\begin{figure}[H]
	\begin{center}
		\includegraphics[scale=.45]{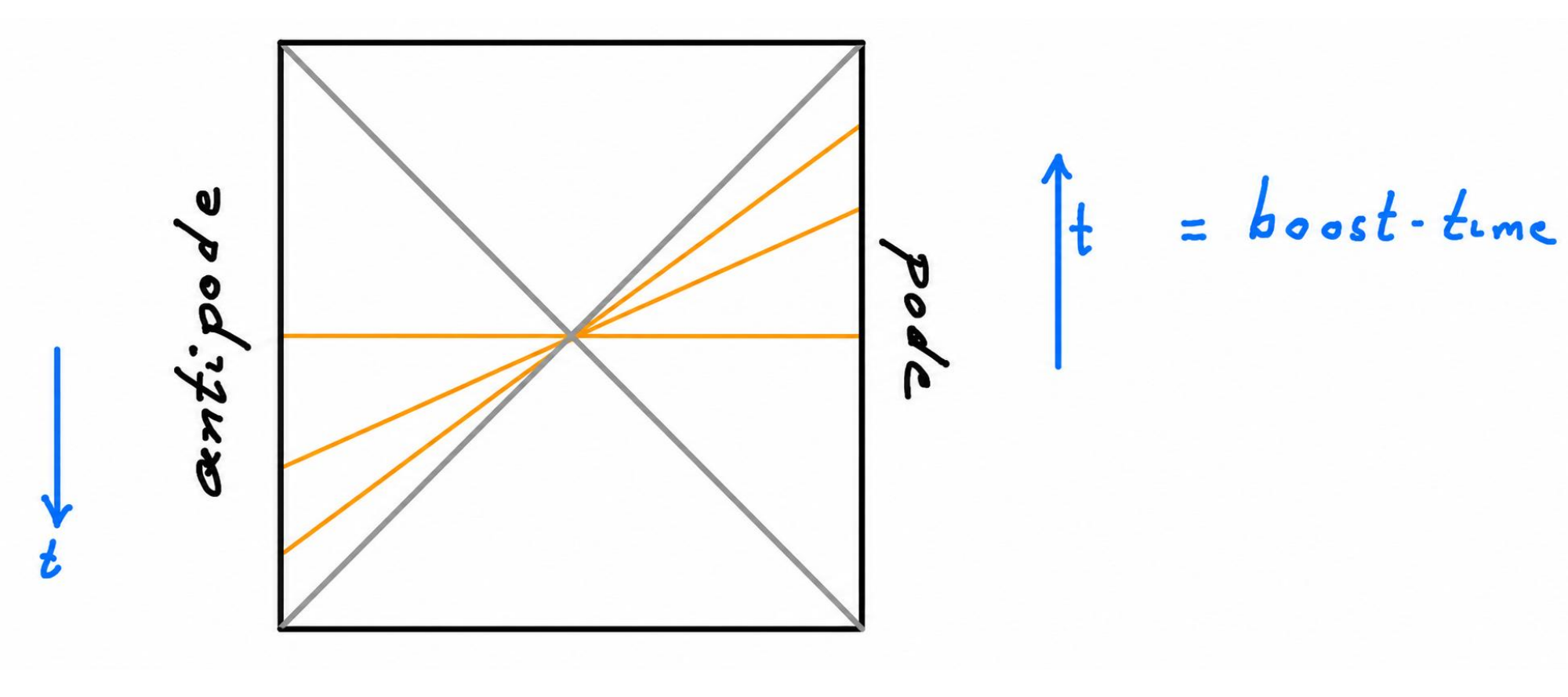}
		\caption{The Penrose diagram for JT-de Sitter: Two opposing static patches; The pode and the antipode; Boost-time.}
		\label{Fig2}
	\end{center}
\end{figure}
The metric and dilaton field of JT-de Sitter in static coordinates are given by
\bea
ds^2 & = & -f(r) dt^2 + \frac{dr^2}{ f(r) } ~ , \\
f(r) & = & \left(1- \frac{r^2}{\ell_{dS}^2} \right) ~ , \\
\phi & = & r ~ .
\eea
\bea
{\rm Pode, ~ Antipode:} &  ~ &  r=0 \notag \\
{\rm Horizon:} & ~ & r= \ell_{dS} \notag 
\eea
The Holographic degrees of freedom, $\psi_{i}$ are in the (stretched) horizon \cite{Banks:2006rx, Susskind:2021dfc}. We assume them to be the fermions in Sachdev--Ye--Kitaev (SYK) model.
\begin{figure}[H]
	\begin{center}
		\includegraphics[scale=.4]{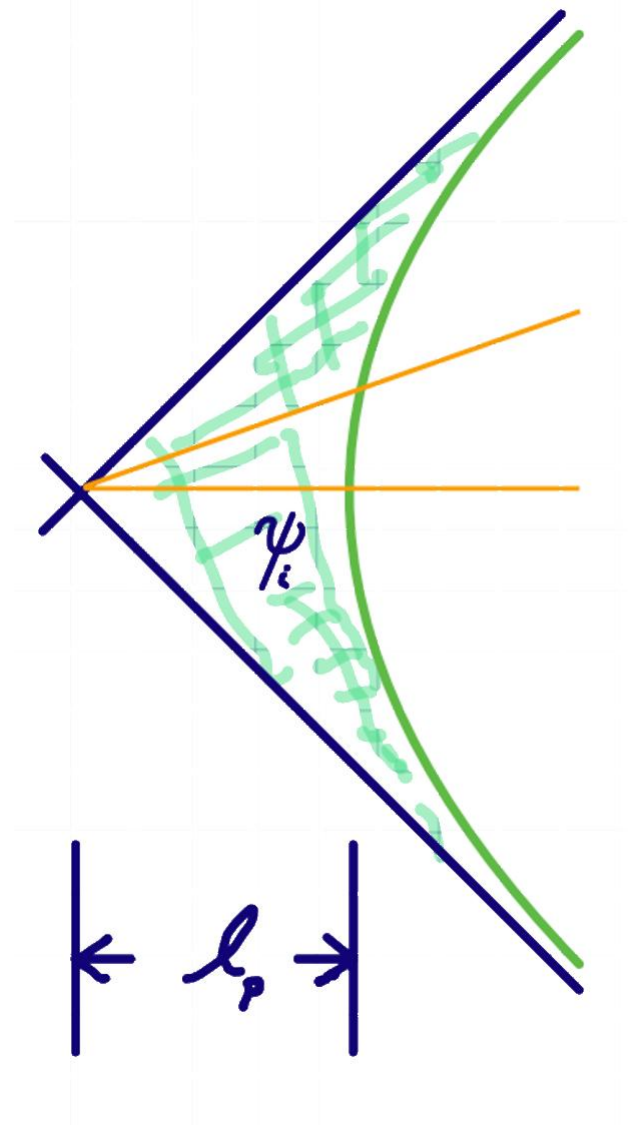}
		\caption{Stretched horizon.}
		\label{Fig3}
	\end{center}
\end{figure} 
~\\
Think of the stretched horizon as a thin layer within a Planck distance $\ell_{P}$ \cite{Susskind:2026vko} from the mathematical horizon at $r= \ell_{dS}$. The Hamiltonian generates ``boosts,''
\be
H = i \partial_{t} ~ .
\ee
\begin{figure}[H]
	\begin{center}
		\includegraphics[scale=.4]{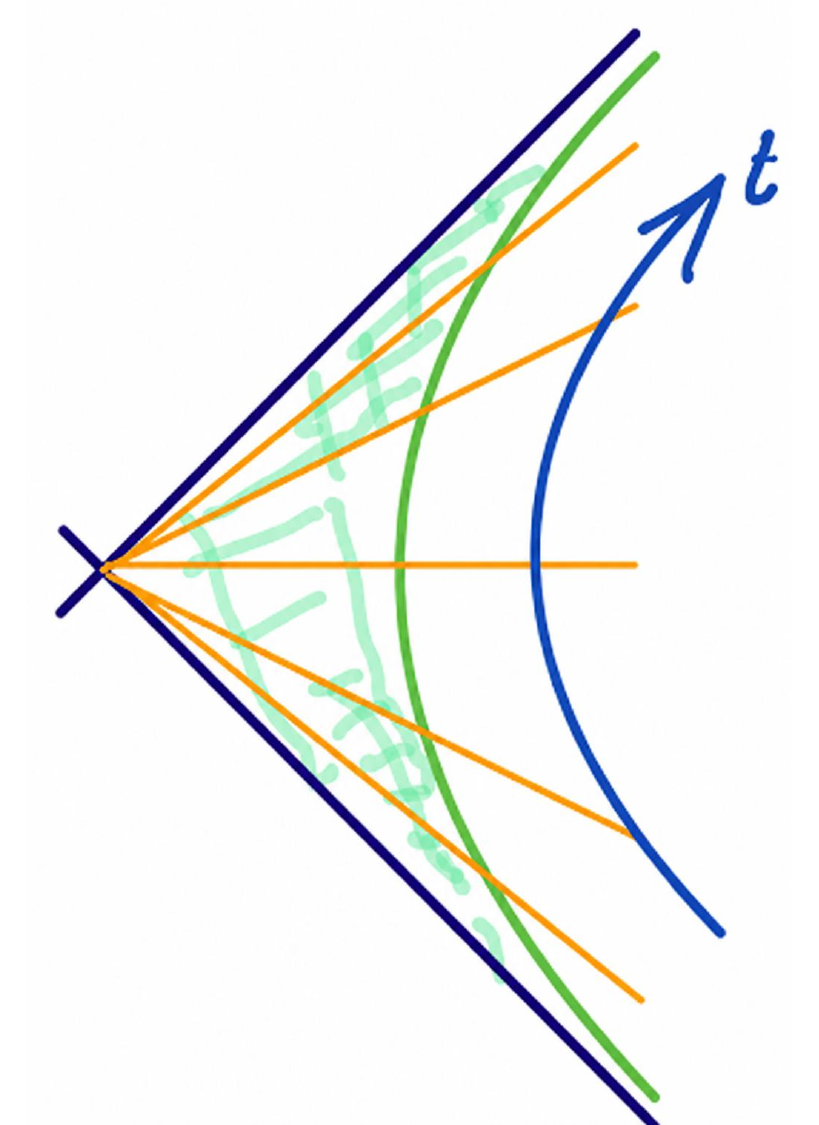}
		\caption{The orange lines are time-slices, time being boost time or static patch time.}
		\label{Fig4}
	\end{center}
\end{figure} 

\subsection{A closer look at the static patch}
\begin{figure}[H]
	\begin{center}
		\includegraphics[scale=.4]{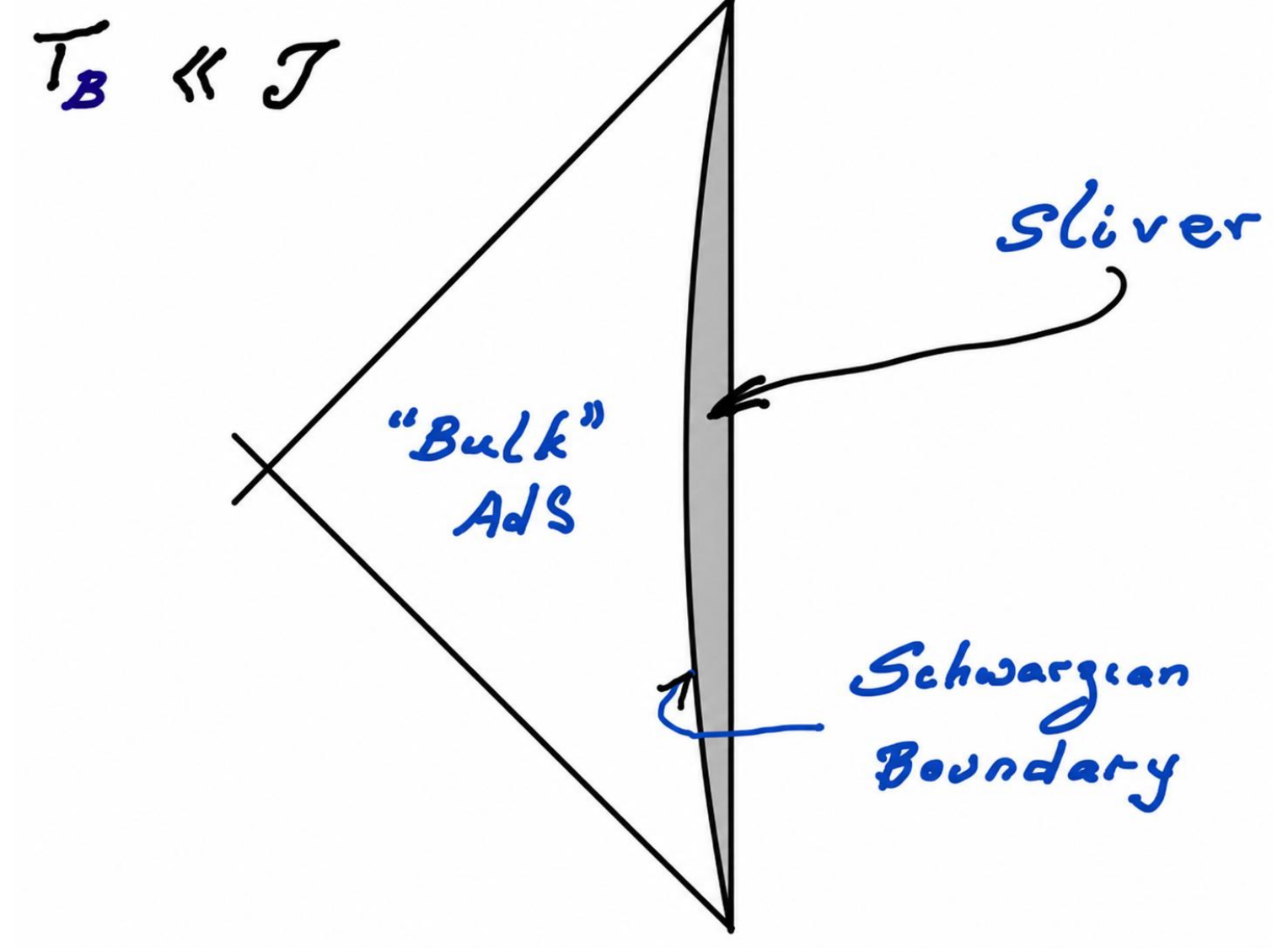}
		\caption{JT static patch. $\mathcal{J}$ is the usual SYK energy scale. The Schwarzian Boundary separates the static patch into bulk AdS and the ``sliver.'' Is there ``life'' in the sliver? }
		\label{Fig5}
	\end{center}
\end{figure}

Let us look more closely at the static patch for JT gravity \cite{Rahman:2024vyg}, first for very low Boltzmann temperature%
\footnote{The Boltzmann temperature $T_{B}$ is the temperature in the thermal density matrix of the Holographic theory
\be
{\rm rho} = \frac{1}{Z} e^{-\frac{H}{T_{B}}} ~.
\ee
It is not the Gibbons--Hawking temperature \cite{Lin:2022nss}.
}
 $T_{B}$.
The overall energy scale of SYK is denoted by a constant $\mathcal{J}$ with units of energy. For $T_{B} \ll \mathcal{J}$, the JT static patch looks like Figure \ref{Fig5}.
As $T_{B}$ increases the Schwarzian surface moves toward the horizon and the sliver grows! The bulk shrinks.
\begin{figure}[H]
	\begin{center}
		\includegraphics[scale=.4]{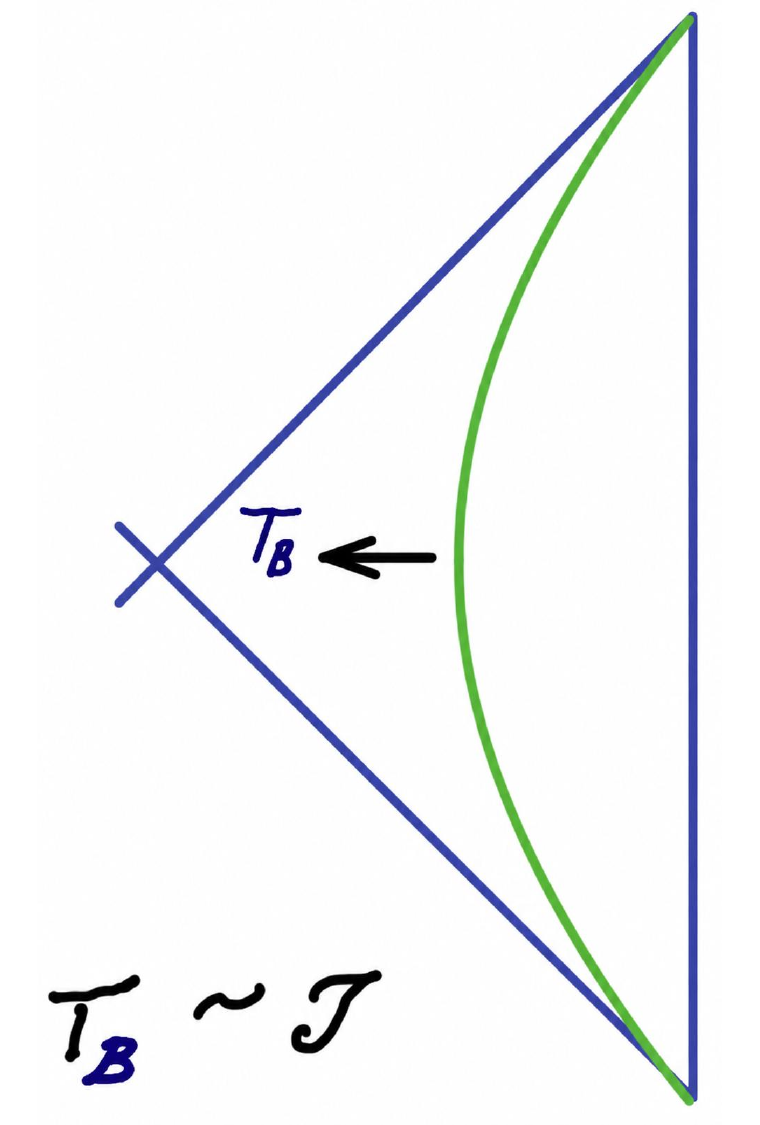}
		\caption{As $T_{B}$ grows the sliver grows. }
		\label{Fig6}
	\end{center}
\end{figure}
Eventually the bulk disappears leaving only the sliver with JT-de Sitter geometry. 
\begin{figure}[h]
	\begin{center}
		\includegraphics[scale=.35]{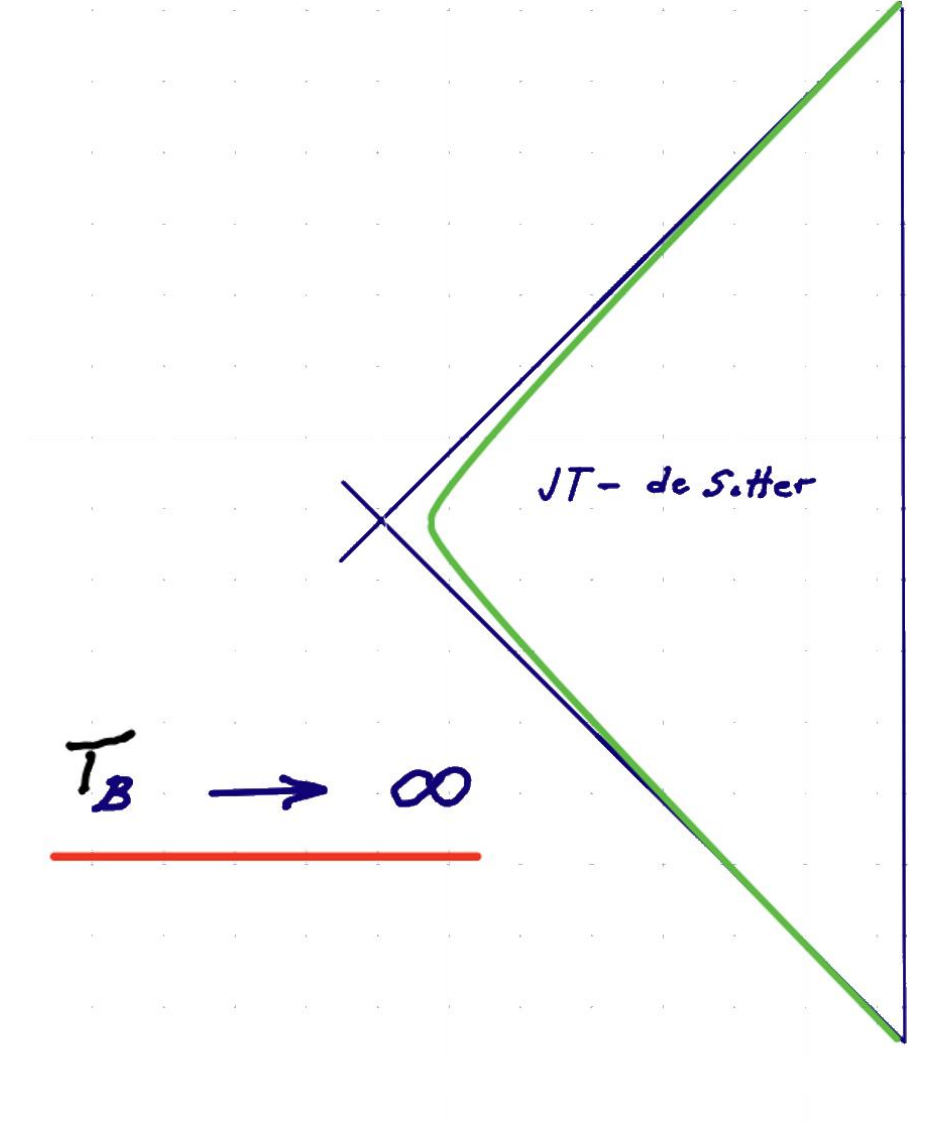}
		\caption{As $T_{B} \to \infty$ the sliver expands to fill the static patch while the original bulk shrinks and becomes the stretched horizon. }
		\label{Fig7}
	\end{center}
\end{figure} 
~\\
\textcolor{red}{\underline{\textcolor{black}{The duality conjecture:}}} \\
\dk ``singlet'' correlations match JT-de Sitter correlations in what was originally the sliver but is now the bulk.
\begin{figure}[H]
	\begin{center}
		\includegraphics[scale=.35]{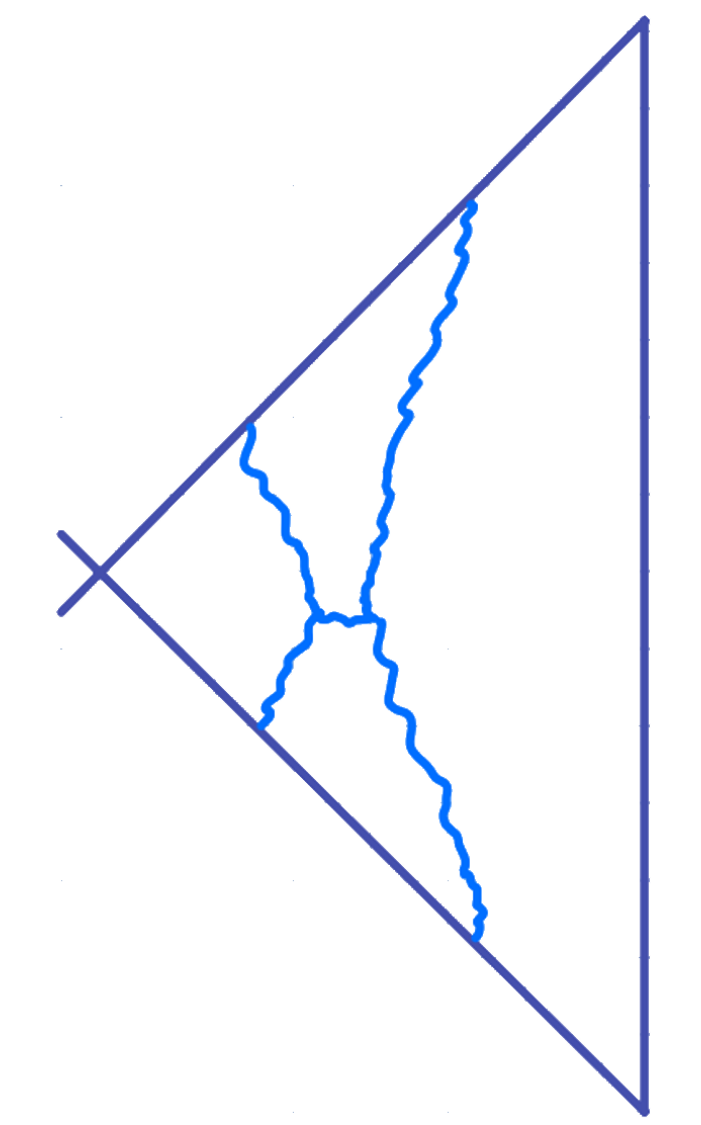}
		\caption{\dk ~ correlator as JT-de Sitter correlator. }
		\label{Fig8}
	\end{center}
\end{figure}
%
%
The duality \dk =JT-dS is unproved but we will assume it. 

\subsection{The Flat-Space Limit}
\begin{figure}[H]
	\begin{center}
		\includegraphics[scale=.4]{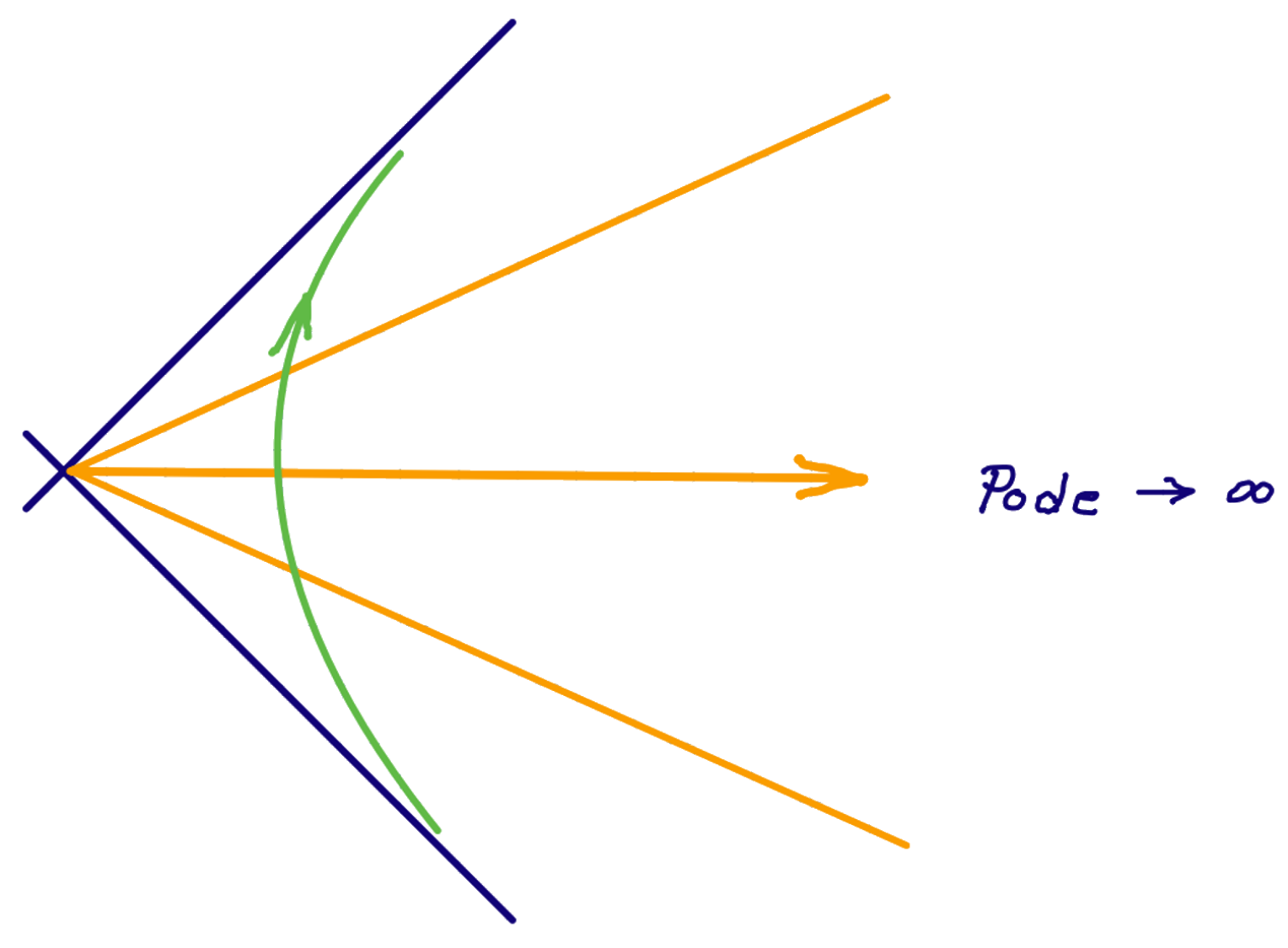}
		\caption{In the flat-space limit, the static patch becomes flat Rindler space.}
		\label{Fig9}
	\end{center}
\end{figure}

In the limit $ \frac{\ell_{dS} }{ \ell_{P}} \to \infty $ with energies kept fixed in Planck units the static patch becomes flat Rindler space. See figure \ref{Fig9}.

\section{DSSYK$_{\infty}$}
Now let's turn to {\textcolor{red}{\underline{\textcolor{black}{\dk}}}}, a theory of $N$ fermion degrees of freedom. There are two cases -- Majorana or real fermions and Dirac or complex fermions. The notations for complex fermions get messy (too many indices) so we will write the equations for the real case, 
\be
\psi_{i} ~~~~~ i= 1,2, \cdots, N ~~~~~ {\rm Majorana} ~ ,
\ee
\be
H = i^{p/2} \sum_{i_{1}<i_{2}<\cdots< i_{p}} J_{i_{1} i_{2} \cdots i_{p}} \psi_{i_{1}} \psi_{i_{2}} \cdots \psi_{i_{p}} ~ ,
\ee
\be
{\rm rho} = e^{-\frac{H}{T_{B}}} = {\rm Density ~ matrix} ~ .
\ee

\subsection{Flat-Space Limit in DSSYK}
The flat-space limit (FSL) is represented by the following limit in SYK;
\begin{itemize}
\item $T_{B} \to \infty$ \\
This is the limit in which the sliver expands to fill the static patch with JT-de Sitter space. For $T_{B}\to \infty$ expectation values are given by normalized trace
\be
\la W \ra = {\rm Tr} (W) ~ .
\ee
\item $N \to \infty$ \\
In the $N\to \infty$ limit $G\to 0$, the entropy $\to \infty$ and the de Sitter radius goes to infinity in Planck units. If we also hold energies fixed this limit is the flat-space limit. Gravity becomes unimportant but the SM remains holographic. 
\item $\frac{p^2}{N} = \lambda \sim 1$ \\
The double-scaled limit. There is an emergent ``string scale'' \cite{Susskind:2022bia}. 
\be
\frac{\ell_{string}}{\ell_{dS}} = \frac{1}{p} \to 0 ~ .
\ee
For fixed $p$ the string scale is parametrically of order the cosmic scale $\ell_{dS}$, but in the double-scaled limit the string theory becomes {\textcolor{red}{\underline{\textcolor{black}{local}}}} on the cosmological scale.
\end{itemize}
In the flat-space limit \cite{Miyashita:2025rpt} the string coupling constant $\bar{g}$ is related to the DSSYK parameter $\lambda = \frac{p^2}{N}$,
\be
\lambda = \bar{g}^2 ~ .
\ee
$\bar{g}$ is the dimensionless open-string splitting and joining coupling constant. We will explain this later.
\begin{figure}[H]
	\begin{center}
		\includegraphics[scale=.4]{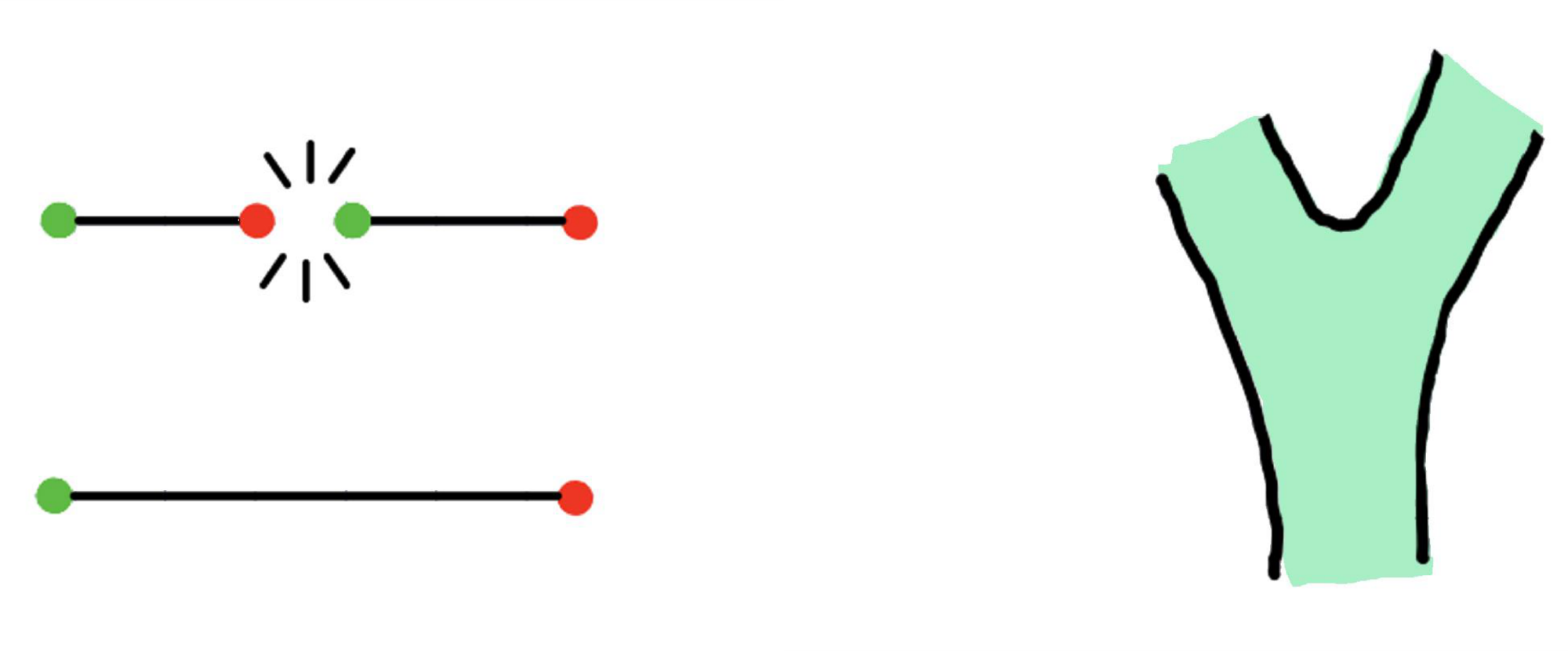}
		\caption{Open string splitting and joining. The coupling constant is $\bar{g}$.}
		\label{Fig10}
	\end{center}
\end{figure}

\subsection{The Ensemble}
The input is the collection of numerical constants
\be
 J_{i_{1} i_{2} \cdots i_{p}} \notag 
\ee
drawn independently from a Gaussian random ensemble,
\be
\mu(J) \sim e^{- \frac{\Sigma J_{i_{1} i_{2} \cdots i_{p}} J_{i_{1} i_{2} \cdots i_{p}}  }{\rm Var(J)}} ~~~~~ {\rm (measure)} ~ .
\ee
The variance is given by 
\be
{\rm Var}(J) = \mathcal{J}^2 \frac{p!}{N^{p-1}} ~~~~~ (\mathcal{J} = {\rm const.}) ~ .
\ee
The dimensionful constant $\mathcal{J}$ is the overall energy scale. (Note: $\mu(J)$ is $O(N)$ or $U(N)$ invariant.)

Consider a general operator:
\be
W(\psi, J) \notag 
\ee
which parametrically depends on $J$.
\bea
\la W \ra & = & {\rm quantum ~ expectation ~ value} \\
~ & = & {\rm Tr} W ~ .
\eea
Typically $\la W \ra$ depends on $J$. 

Ensemble averages are denoted by a bar:
\be
\overline{f} = \int_{J} \mu(J) f(J) ~ .
\ee
An example: The ensemble average of the expectation value,
\be
\overline{\la W \ra} = \int_{J} \mu(J) {\rm Tr} W ~ .
\ee

\subsection{Ensemble Averaged (EA) Theory}

The EA theory is defined by integrating over the (time independent) parameters $ J_{i_{1} i_{2} \cdots i_{p}}$ as if they were degrees of freedom, 
\bea
Z = \int dJ {\rm Tr} \exp\Big\{- i^{p/2} \sum   J_{i_{1} i_{2} \cdots i_{p}} \psi_{i_{1}} \psi_{i_{2}} \cdots \psi_{i_{p}} 
+ \frac{\sum  J_{i_{1} i_{2} \cdots i_{p}}  J_{i_{1} i_{2} \cdots i_{p}} }{ \rm Var(J)} \Big\} ~ . \notag \\
\eea
The EA theory is $O(N)$ or $U(N)$ invariant. SYK calculates the EA theory. That's not our goal but let's proceed.

\subsection{SYMMETRY of the EA Theory}

Let us consider the following transformations of fermions:
\bea
O(N): & \psi_{\ell}^{\prime} = O_{\ell i} \psi_{i} & ~~~~~~~~ {\rm Majorana} ~ , \\
~ & {\rm or} & ~ \notag \\
U(N): & \psi_{\ell}^{\prime} = U_{\ell i} \psi_{i} & ~~~~~~~~ {\rm Dirac} ~ .
\eea
In order to define the symmetry $O(N)$ must also act on $J_{i_{1} i_{2} \cdots i_{p}} $, 
\bea
J_{lpq \cdots}^{\prime} = O_{li} O_{pj} O_{qk} \cdots J_{ijk \cdots } ~ , \\
({\rm similar ~ for ~} U(N))  ~ .
\eea
The EA theory is $O(N)$ or $U(N)$ invariant. 

In the {\textcolor{red}{\underline{\textcolor{black}{Unaveraged (UA) theory}}}}, the $J$'s are a fixed set of constants. Observables parametrically depend on $J$. Each instance is \underline{not} invariant under $O(N)$. Our goal is to understand the properties of individual instances of the unaveraged theory in the flat-space limit. To that end we construct diagrammatic Feynman rules \cite{Sekino:2025bsc}.

\subsection{Diagrammatic Rules (for the $O(N)$ theory)}
{\textcolor{red}{\underline{\textcolor{black}{Vertices:}}}}
\be
 J_{i_{1} i_{2} \cdots i_{p}} \psi_{i_{1}} \psi_{i_{2}} \cdots \psi_{i_{p}} = \raisebox{-1.2cm}{ \includegraphics[scale=.2]{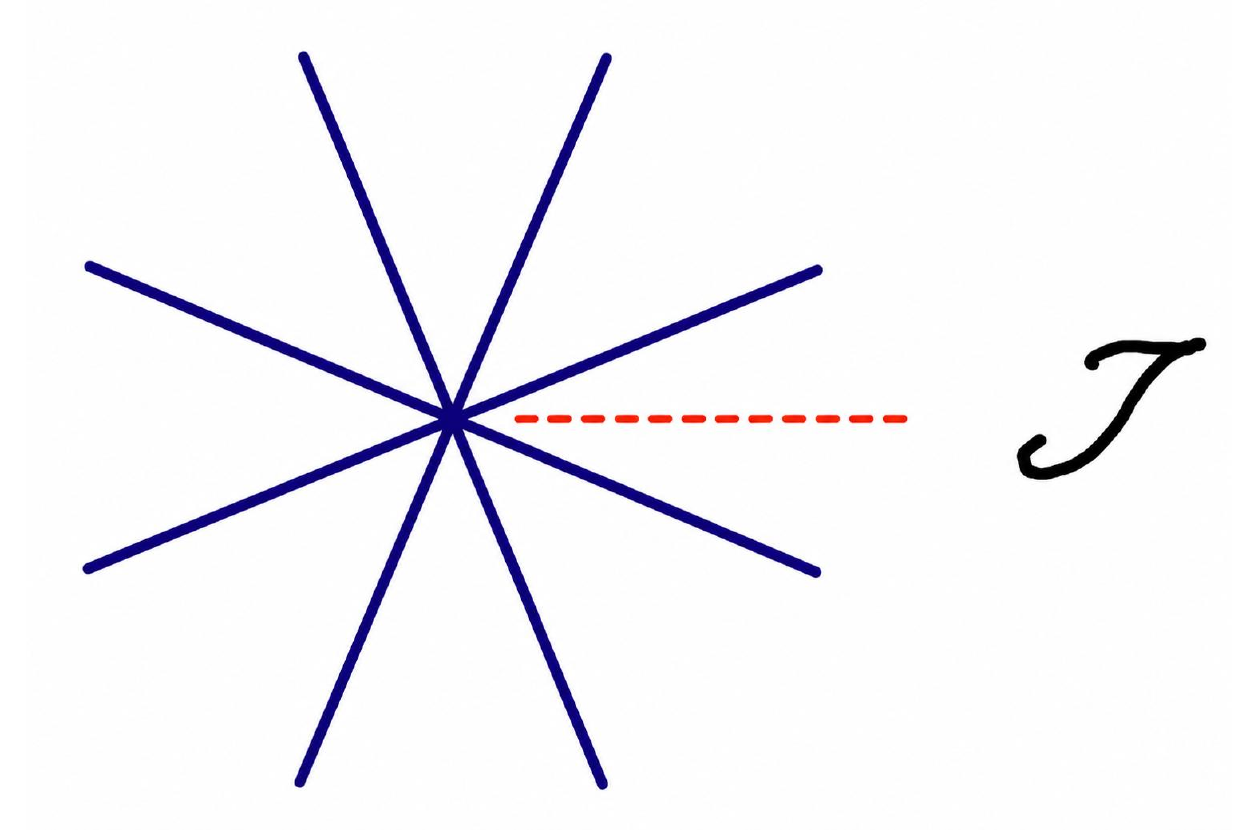}} ~ .
\ee
Black solid lines represent fermions. The vertex represents monomials $\psi_{i_{1}} \psi_{i_{2}} \cdots \psi_{i_{p}} $. The dashed red lines represent the factors $J_{i_{1} i_{2} \cdots i_{p}} $. Each vertex is assigned a numerical weight $\mathcal{J}$. \\
{\textcolor{red}{\underline{\textcolor{black}{Propagators:}}}}\\
~\\
Fermion propagators
\be
\raisebox{-0.6cm}{ \includegraphics[scale=.2]{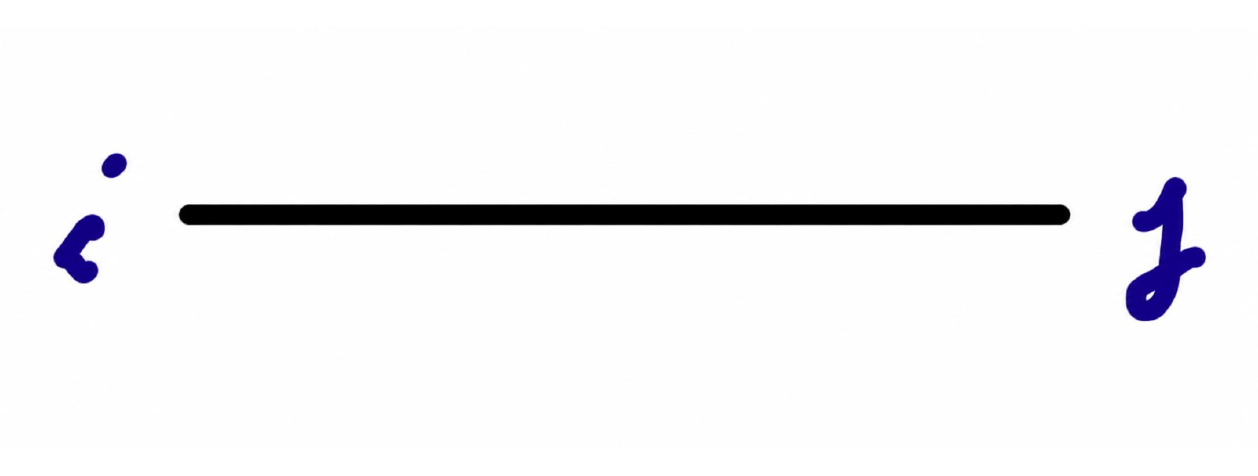}} = \epsilon (t-t') \delta_{ij} ~ .
\ee
$\overline{JJ}$ propagators
\be
\raisebox{-0.6cm}{ \includegraphics[scale=.2]{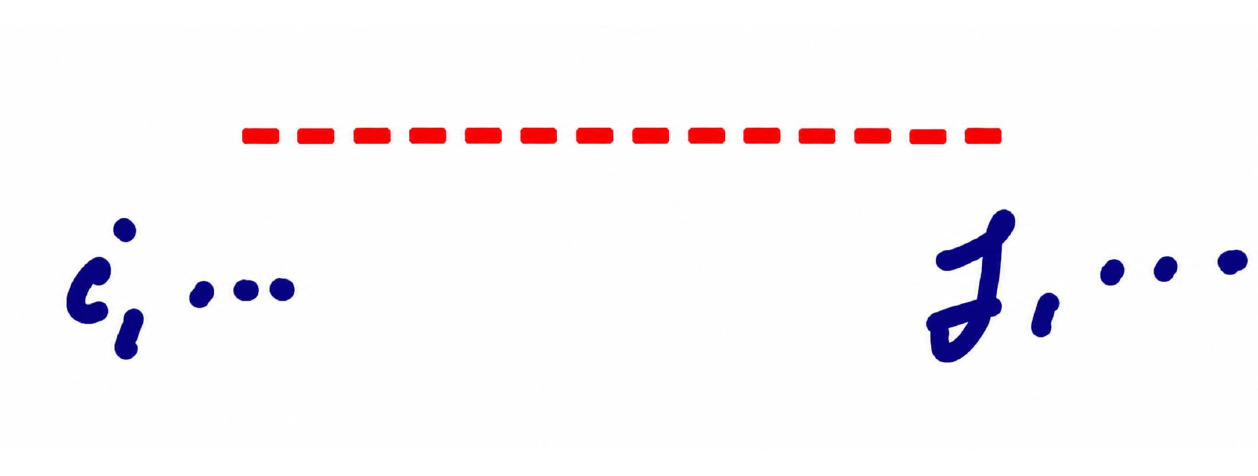}} = \frac{p!}{N^{p-1}} \delta_{i_{1} i_{2} \cdots, j_{1}j_{2} \cdots} ~ .
\ee

\begin{figure}[h]
	\begin{center}
		\includegraphics[scale=.3]{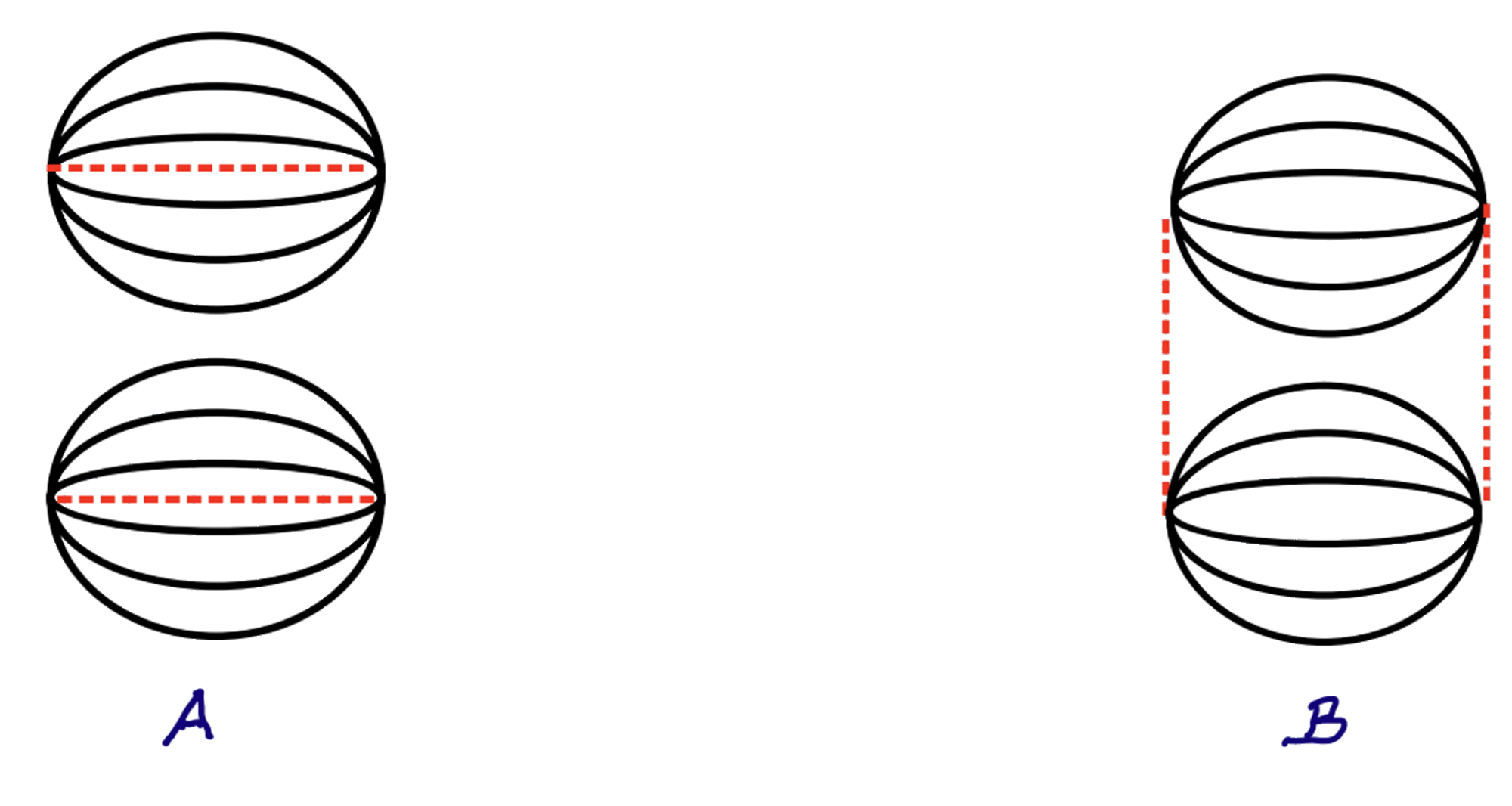}
		\caption{{\textcolor{red}{\underline{\textcolor{black}{Diagrams A and B.}}}}}
		\label{Fig12}
	\end{center}
\end{figure}

Fig. \ref{Fig12} are two diagrams that differ only in the placement of the \raisebox{-0.03cm}{ \includegraphics[scale=.1]{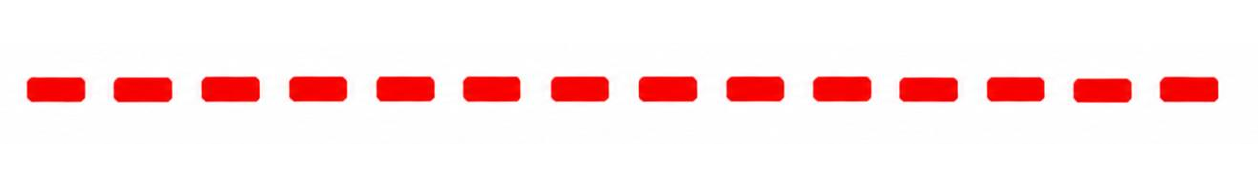}}  propagators. 
 \\
In Diagram A the red propagators are fully ``accompanied'' by fermions. This means that each red line has a full set of $p$ fermion lines alongside it going between the same endpoints. 

In diagram B there are no fermions accompanying the \raisebox{-0.03cm}{ \includegraphics[scale=.1]{FigEqPropJJ2}}  propagators. Both diagrams have two \raisebox{-0.03cm}{ \includegraphics[scale=.1]{FigEqPropJJ2}} lines giving a factor
\be
\frac{p!}{N^{p-1}} \frac{p!}{N^{p-1}} \notag ~ .
\ee
There are also combinatoric factors which count the numbers of ways of choosing the fermions. For diagram A the combinatoric factor is $\frac{N^{p}}{p!} \frac{N^{p}}{p!} $ but by carefully tracking the indices one finds that for diagram B there is only one factor of $\frac{N^{p}}{p!} $.

Combining the factors;
\bea
{\rm A}: & \displaystyle \frac{p!}{N^{p-1}} \frac{N^{p}}{p!} \frac{p!}{N^{p-1}} \frac{N^{p}}{p!} = N^2 \\  
{\rm B}: & \displaystyle \frac{p!}{N^{p-1}} \frac{p!}{N^{p-1}} \frac{N^{p}}{p!} =  \frac{p!}{N^{p-2}} 
\eea
B (relative to A) is not only smaller than any power of $N$ in the double-scaled limit: It is smaller than any exponential in $N^{1/2}$, 
\be
\frac{{\rm B}}{\rm A} < {\textcolor{red}{\underline{\textcolor{black}{e^{-a N^{1/2}}}}}} ~~~~~~~ ({\rm all} ~ a) ~ .
\ee

Now let's return to a general operator
\be
W = W(\psi, J) ~ . \notag 
\ee
Diagrammatically,
\be
W(\psi, J) = \raisebox{-1.2cm}{ \includegraphics[scale=.2]{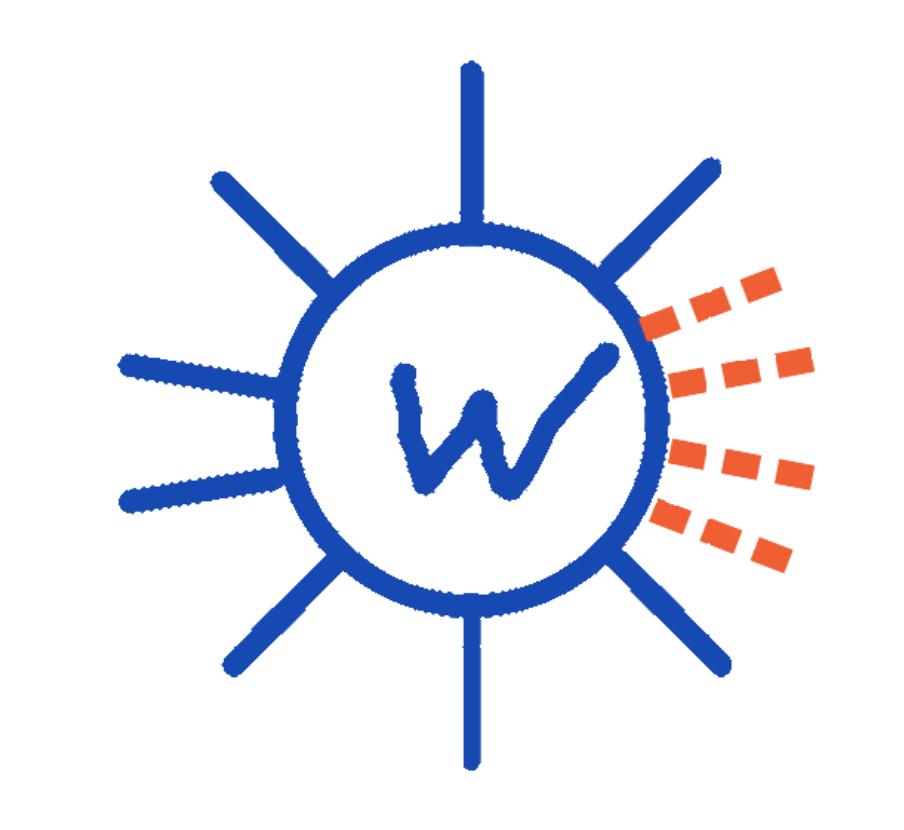}} ~ .
\ee
The general rule for taking expectation values is to contract fermion lines in all possible ways. The rule for ensemble averaging is to contract the red lines in all possible ways. To take the ensemble average of an expectation value we contract all lines. 

For simplicity let $\overline{\la W \ra} = 0$. (There is no loss of generality in setting $\overline{\la W \ra} = 0$. If it's not zero we can subtract a constant.) Thus --
\be
\overline{ \la W \ra} = \raisebox{-1.2cm}{ \includegraphics[scale=.2]{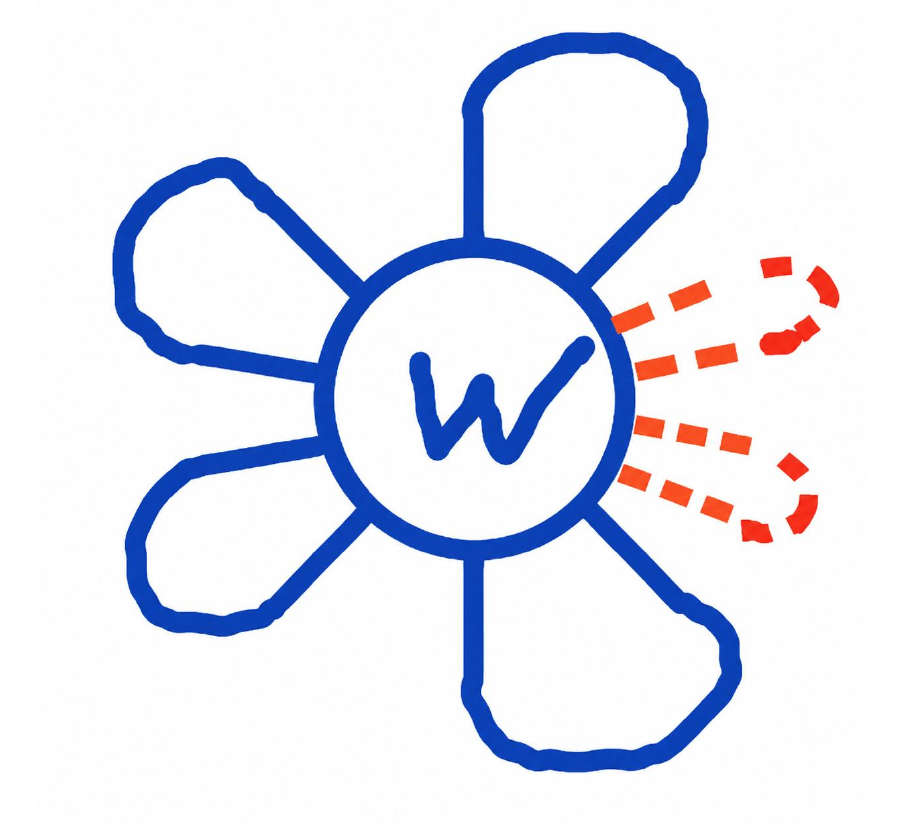}} = 0 ~ .
\ee
In general $\la W \ra \neq 0$ (No EA). It is a function of $J$  
\be
 \la W \ra = \raisebox{-1.2cm}{ \includegraphics[scale=.2]{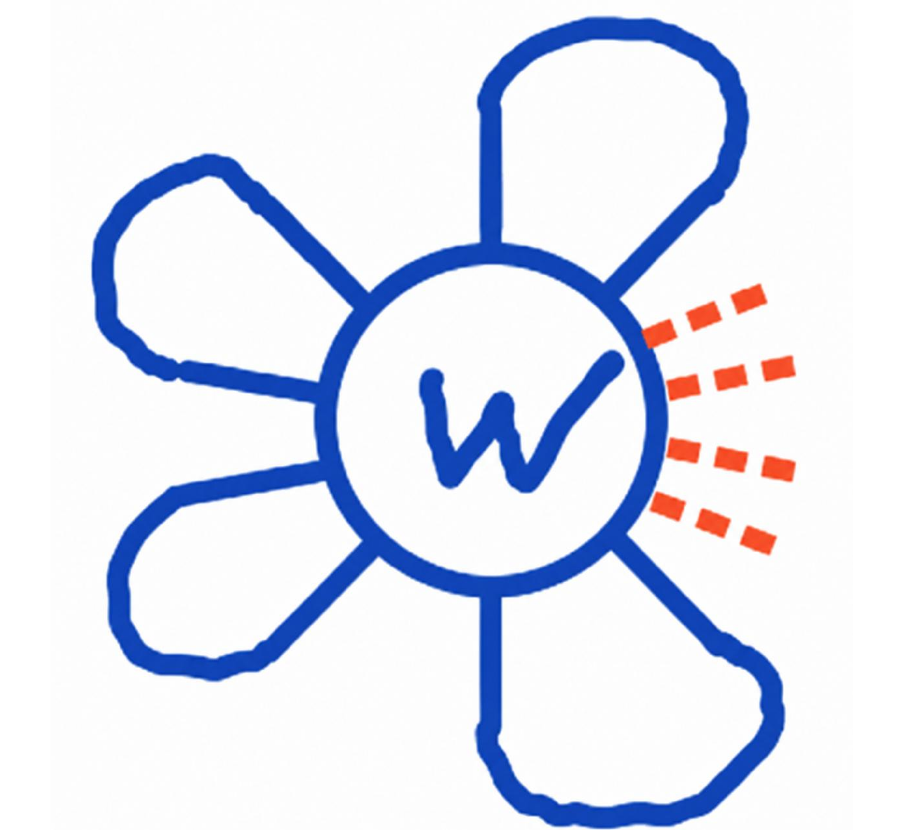}} \neq 0 ~ .
\ee
We would like to know to what extent the ``self-averaging'' occurs. ${\rm Var} \la W \ra$ is a measure of how far $\la W \ra$ deviates from $\overline{ \la W \ra}$ as $J$ is varied. Self-averaging simply means that ${\rm Var} \la W \ra$ is small. How small we will see. Let's therefore calculate 
\be
{\rm Var} \la W \ra = \overline{\la W \ra \la W \ra } ~ .\notag \\
\ee
We begin with $\la W \ra \la W \ra$ not ensemble averaged. 
\be
\la W \ra \la W \ra = \raisebox{-1.2cm}{ \includegraphics[scale=.2]{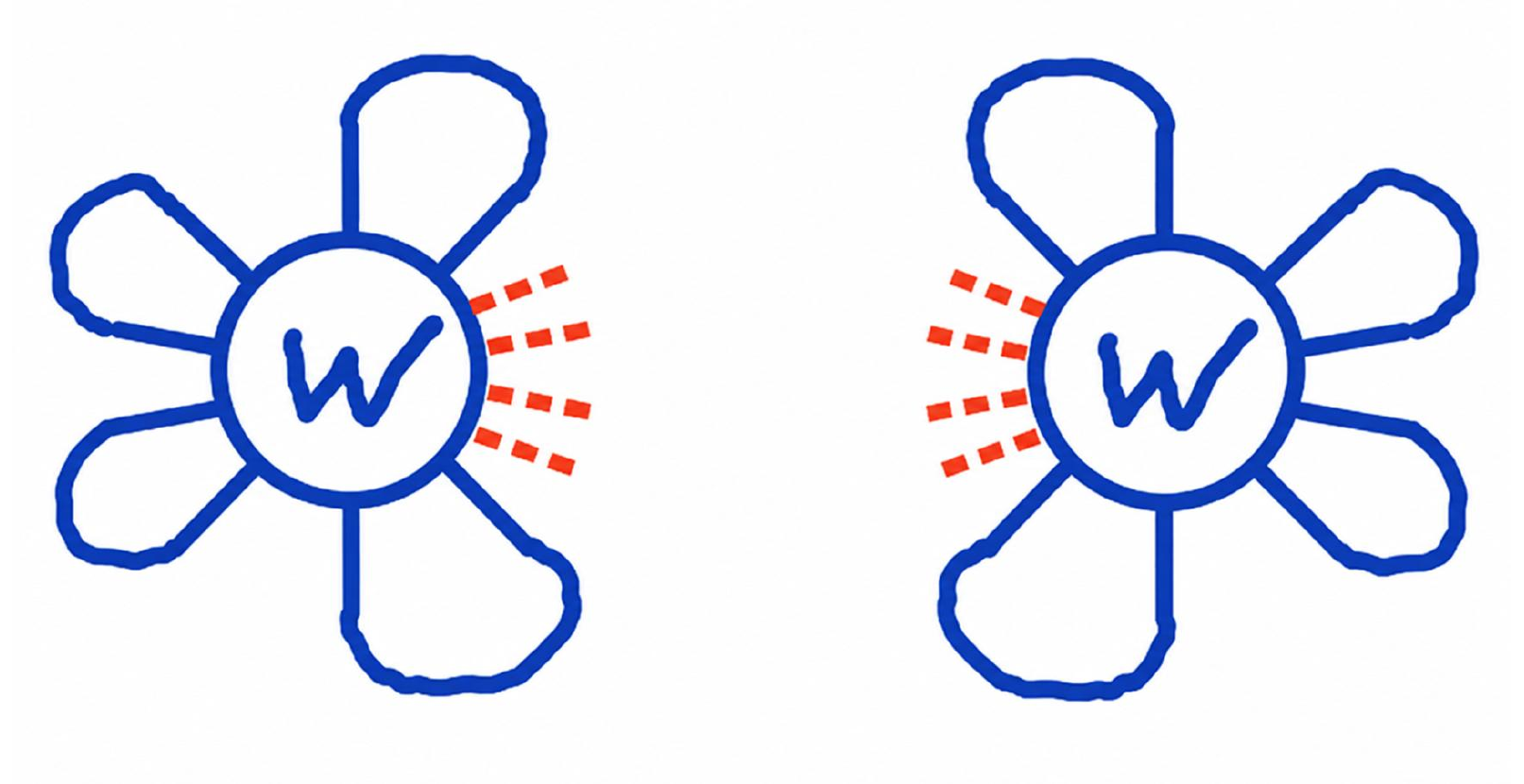}} . \notag 
\ee
To ensemble average ${\rm Var} \la W \ra$ we contract the red lines in all ways. 
\be
\overline{ \la W \ra \la W \ra }= \raisebox{-1.2cm}{ \includegraphics[scale=.2]{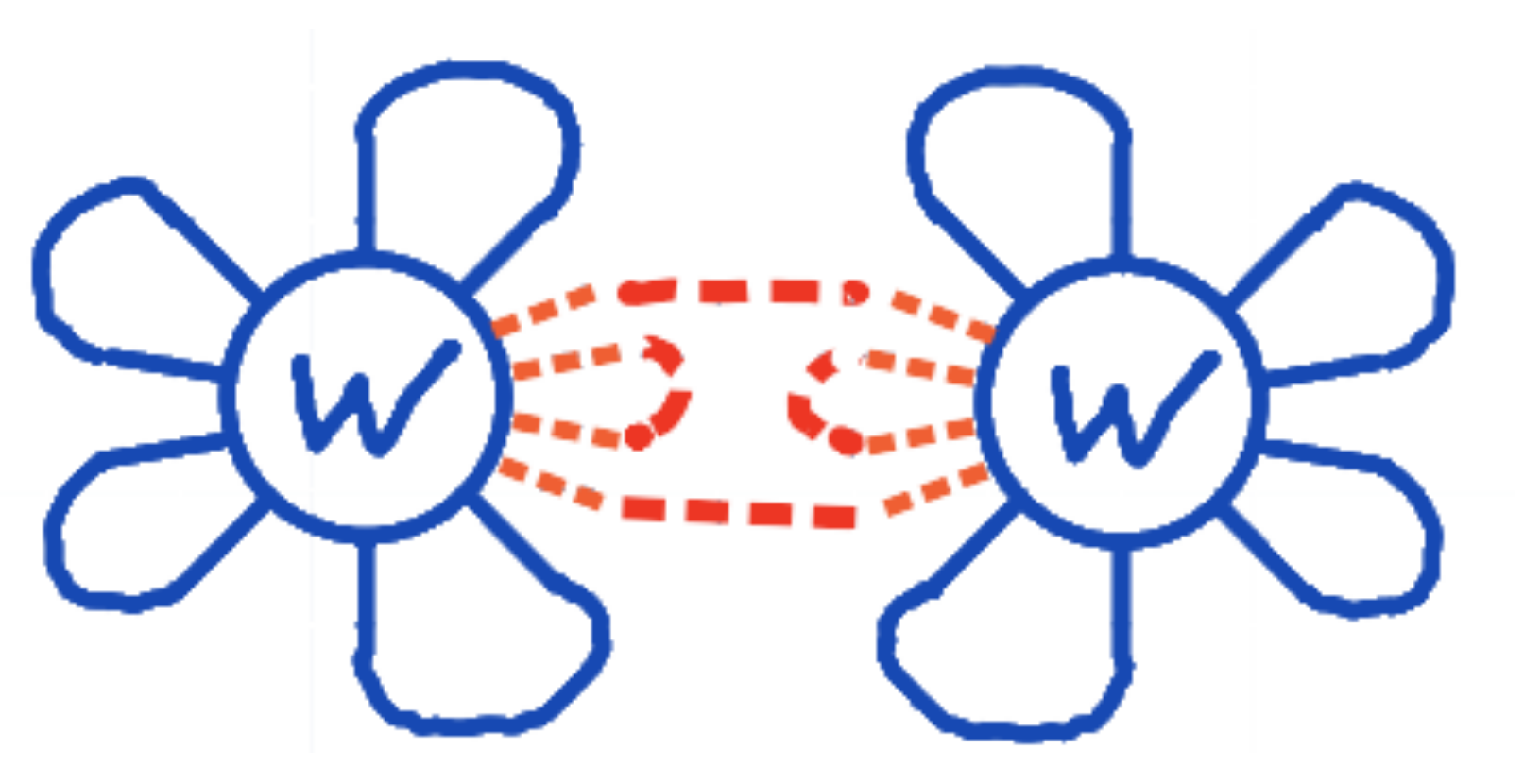}} . \notag 
\ee
Recall the diagram B of Figure \ref{Fig12}: there are no fermions accompanying red dotted lines.   
One finds
\bea
{\rm Var} \la W \ra & = & \overline{ \la W \ra \la W \ra} \approx \frac{p!}{N^{p}} \notag \\
 ~ & < & e^{-a \sqrt{N}} ~~~~~~~ ({\rm all} ~ a).
\eea
${\rm Var} \la W \ra$ is not only smaller than any power of $N$ but smaller than any exponential in $\sqrt{N}$. We call this 
\begin{center}
{\textcolor{red}{\underline{\textcolor{black}{``Perfect self-averaging.''}}}} 
\end{center}

The EA theory is $O(N)$ (or $U(N)$) invariant. The correlation function between a singlet $\sum\psi_{i}\psi_{i}$ and an adjoint $\psi_{j}\psi_{k}$ vanishes in an $O(N)$ invariant theory. Nonvanishing of such quantities indicates violations of the symmetry. The variance found above suggests that in the UA theory, violations of $O(N)$ are not zero for finite $N$, but they do tend to zero faster than any exponential in $N^{1/2}$.
\bea
      \begin{aligned}
      {\textcolor{red}{\underline{\textcolor{black}{\rm violations}}}} ~~~~ \\
      {\textcolor{red}{\underline{\textcolor{black}{\rm of ~ symmetry}}}}
      \end{aligned}
      ~~~ \leq ~~~
     {\textcolor{red}{\underline{\textcolor{black}{\exp(-aN^{1/2})}}}}  \\
           {\textcolor{red}{\underline{\textcolor{black}{ {\rm all} ~ a }}}} \notag 
\eea
Let's now consider the flat-space limit,
\bea
\frac{\ell_{dS}}{\ell_{P}}  & \to & \infty ~ , \notag \\
G & \to & 0 ~, \\
N & \to & \infty ~ . \notag 
\eea
This is precisely the limit in which self-averaging becomes perfect. Therefore we may assume that in the FSL $O(N)$ or $U(N)$ symmetry is exact.

\section{Gauge Theory}
What are the implications? Recall: Symmetry of a hologram becomes gauge symmetries of the bulk. The $O(N)$ or $U(N)$ symmetry of the hologram implies that the bulk dual is a gauge theory. \\
~ \\
~~~ \fcolorbox{red}{white}{ \begin{minipage}{14cm} The SM of \dk in the flat-space limit is an $O(N)$ (or $U(N)$) gauge theory. \\
(Expressed in Rindler coordinates; recall Fig. \ref{Fig9}.) \end{minipage}}\\
~\\

An obvious candidate is the 't~Hooft model (1+1 dimensional QCD).

\subsection{Gauge theory \dk ~ Correspondence \cite{Miyashita:2025rpt}}
Let us compare parameters of the two theories. \\
{\textcolor{red}{\underline{\textcolor{black}{Parameters in the 't~Hooft Model:}}}} 
\begin{enumerate}
\item $N_{ym}$ as in the Yang-Mills gauge group $O(N_{ym})$ or $U(N_{ym})$. 
\item $g$, the gauge coupling with units of mass. 
\item $m_{q}^2$, the renormalized quark squared mass. The bare and renormalized mass satisfy
\be
m_{q}^2 = m_{bare}^2 - \frac{g^2 N_{ym}}{\pi} ~ .
\ee
\item $\displaystyle \overline{g}^2  =  \frac{g^2}{m_{q}^2}, $ dimensionless gauge coupling and string coupling. 
\item $\alpha = g^2 N_{ym}$, the 't~Hooft coupling usually called $\lambda$. To avoid confusion we call it $\alpha$. 
\item $\overline{\alpha} = \overline{g}^2 N_{ym}$, dimensionless 't~Hooft coupling.
\end{enumerate}
{\textcolor{red}{\underline{\textcolor{black}{Parameters in \dk:}}}} 
\begin{enumerate}
\item $N_{syk}$, the number of SYK fermions.
\item $p$, the order of the monomials $\psi_{i_{1}} \psi_{i_{2}} \cdots \psi_{i_{p}} $. 
\item $\lambda = p^2/N_{syk}. $
\end{enumerate}
The correspondences \cite{Miyashita:2025rpt} are
\begin{itemize}
\item $N_{ym} = N_{syk}$. \\ From now on we will denote them both by $N$.
\item $\overline{\alpha} = p^2$. 
\item $\overline{g}^2 = \lambda$. 
\end{itemize}
Note the similarity of the defining equations for $\overline{\alpha}$ and $p^2$,
\bea
\overline{\alpha} & = & \overline{g}^2 N ~ , \\
p^2 & = & \lambda N ~ .
\eea
The 't~Hooft limit 
\be
N\to \infty ~ , ~~~~ \overline{\alpha} = ~ {\rm fixed} \label{tHL}
\ee
parallels the limit
\be
N\to \infty~, ~~~~ p^2 = ~ {\rm fixed} ~ . \label{pL}
\ee
The flat-space limit 
\be
N\to \infty ~ , ~~~~ \overline{g}^2 = ~ {\rm fixed} \label{gL}
\ee
parallels the double-scaled limit 
\be
N\to \infty ~ , ~~~~ \lambda = ~ {\rm fixed} ~ . \label{DSL}
\ee
Having argued that the SM following from \dk ~ is a Yang-Mills gauge theory, what about the matter content? Let's compare the Rindler entropy of a gauge theory in flat space with the entropy of \dk. Typically
\be
S_{gauge} \sim N^2 + S_{matter} ~~~~ ({\rm Gauge ~ fields ~ in ~ adjoint} ) ~ .
\ee
But 
\be
S_{DSSYK_{\infty}} = N \log 2 ~ ,
\ee
is there a mismatch? No. In $D=(1+1)$ gauge bosons are non-dynamical. The gauge fields do play an important role in mediating long range $(1+1)$ dim instantaneous Coulomb forces but the entropy, either of a hot plasma, or Rindler space is entirely due to matter. Therefore all we need for a match is:
\be
{\rm dim ~ matter} = N ~ .
\ee
The obvious choice is matter $=$ fermions, i.e., quarks in the $N$-dimensional fundamental. \\

\fcolorbox{red}{white}{ \begin{minipage}{14cm} 
SM dual of \dk ~ is 't~Hooft model, QCD(2) + QED. 
\be
N_{color}  =  N_{syk} = N ~, 
\ee
\be
N_{flavor}  =  1 ~ .
\ee
\end{minipage}}\\
~\\
In the $U(N)$ theory the SYK fermions correspond to electrically charged quarks. 
\bea
\psi_{i} & \longleftrightarrow & q_{i} \\
\psi_{i}^{\dagger} & \longleftrightarrow & q_{i}^{\dagger}
\eea
The SM of \dk ~ is the 't~Hooft model -- QCD$_{2}$ $\times$ $U(1)$. Let us be more precise. By the 't~Hooft model one usually means the limit $N \to \infty$  with $\alpha= g^2 N$ fixed. That limit does {\textcolor{red}{\underline{\textcolor{black}{not}}}} correspond to the double-scaled limit \eqref{DSL}. 't~Hooft scaling \eqref{tHL} corresponds to \eqref{pL}. The flat-space limit is the limit \eqref{gL}
corresponding to $\lambda$ fixed \eqref{DSL}. However to compare SYK with the 't~Hooft model we need to work in the fixed $\overline{g}^2 N$ limit since that's where the calculations are done. This means $N \to \infty$ with $p^2 = \lambda N$ fixed.

The particle spectrum of the 't~Hooft model consists of two degenerate Regge trajectories, a CP-even trajectory and a CP-odd trajectory. In the limit $\overline{g}^2 N = \overline{\alpha} = {\rm fixed}$ the two trajectories lie on top of one another. The spectrum is computed by summing ladder graphs.

\subsection{'t~Hooft meson spectrum}
\begin{figure}[H]
	\begin{center}
		\includegraphics[scale=.3]{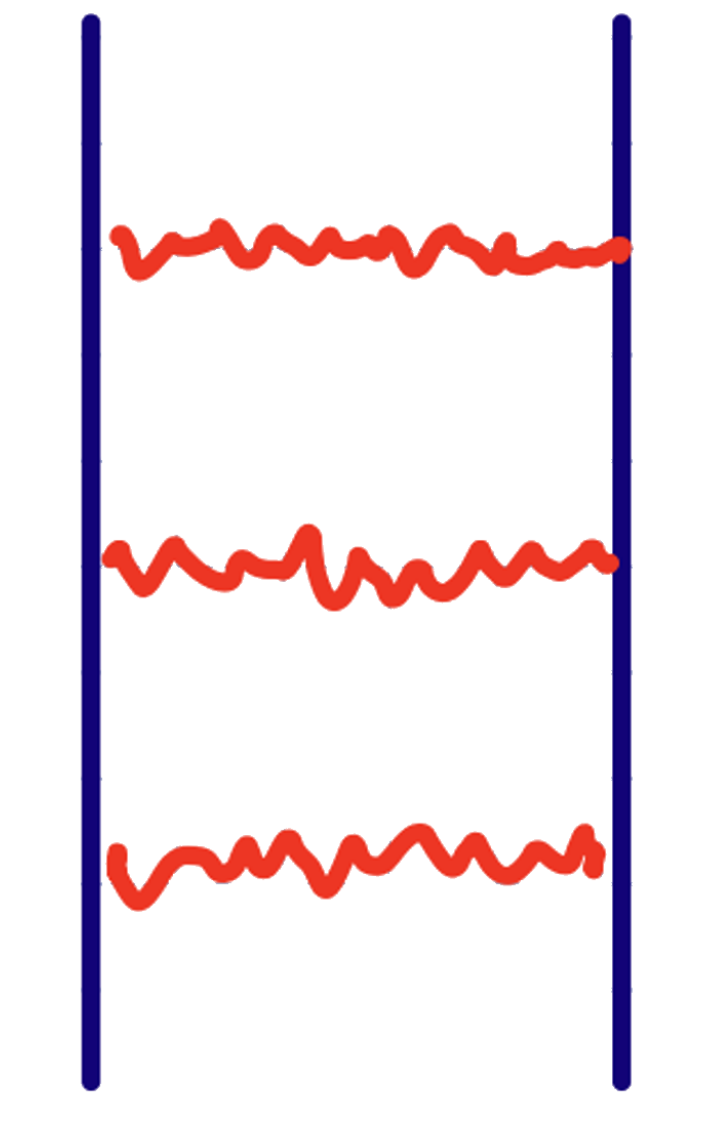}
		\caption{Ladder summation.}
		\label{Fig13}
	\end{center}
\end{figure}
The gauge boson exchanges give rise to a confining linear potential. The potential is due to a uniform chromoelectric flux line or an open string between the quarks. The 't~Hooft model is the simplest of all string theories.

Let's go further with the parameter correspondence. Call the string tension in the 't~Hooft model $\tau$,
\bea
\tau = g^2 N & = & \overline{\alpha} m_{q}^2 \\
~ & = & p^2 m_{q}^2 ~ .
\eea
In the double-scaled limit this is 
\be
\tau = N m_{q}^2 \lambda ~ .
\ee
In QCD$_{2}$ the string tension is kept fixed and finite and in DSSYK $\lambda$ is fixed and finite. This requires the renormalized quark mass to tend to zero
\be
m_{q}^2 \to 0 ~ .
\ee
(Note: The limit $m_{q} \to 0$ is not the chiral limit. In the chiral limit the bare quark mass $\to 0$. The bare and renormalized mass-squared differ by an additive term $g^2 N/\pi$. The theory with $m_{q}^2 = 0$ was analyzed in \cite{Fateev:2009jf}. It is a perfectly regular theory with a nonvanishing mass gap.)

\begin{figure}[H]
	\begin{center}
		\includegraphics[scale=.55]{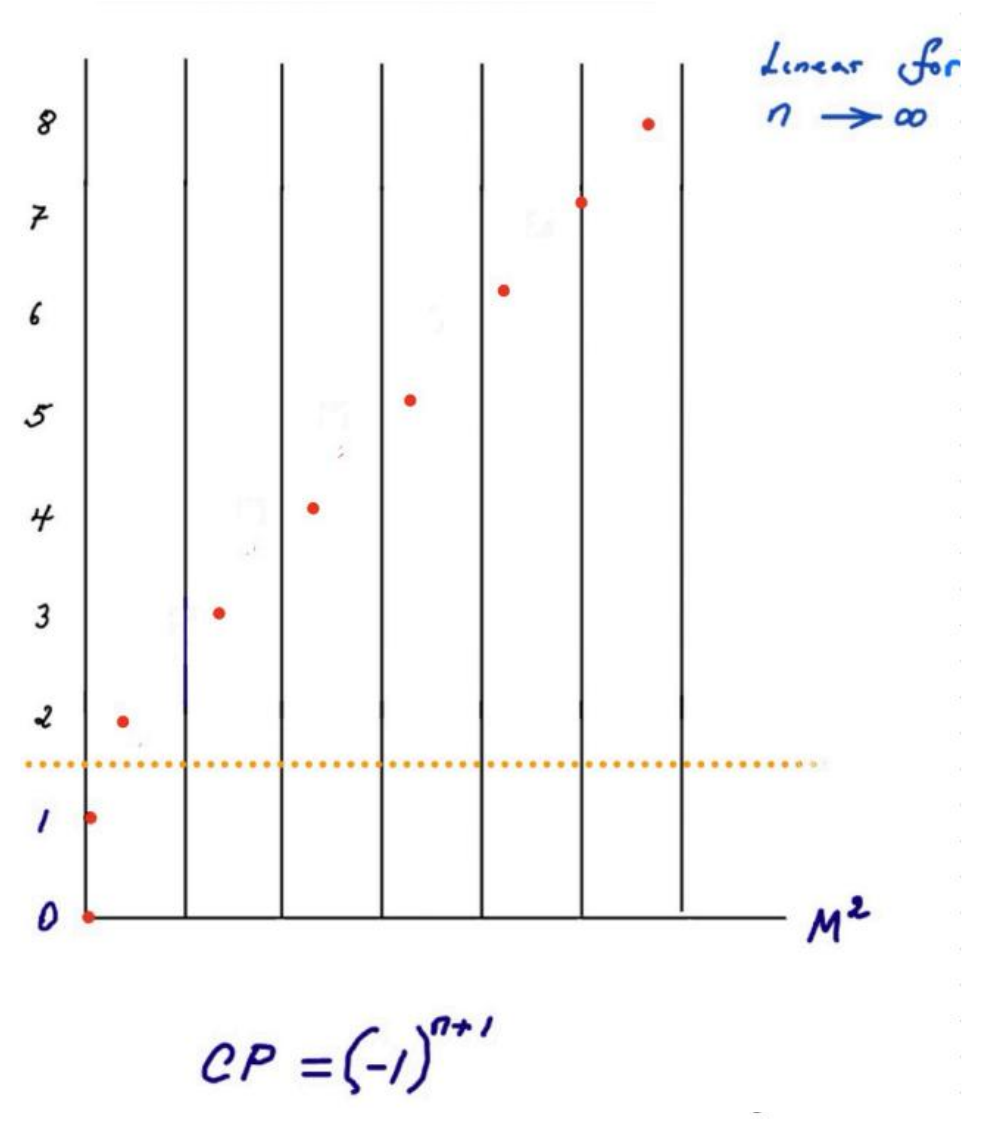}
		\caption{Meson spectrum in the 't~Hooft model.}
		\label{Fig14}
	\end{center}
\end{figure}

\begin{figure}[H]
	\begin{center}
		\includegraphics[scale=.35]{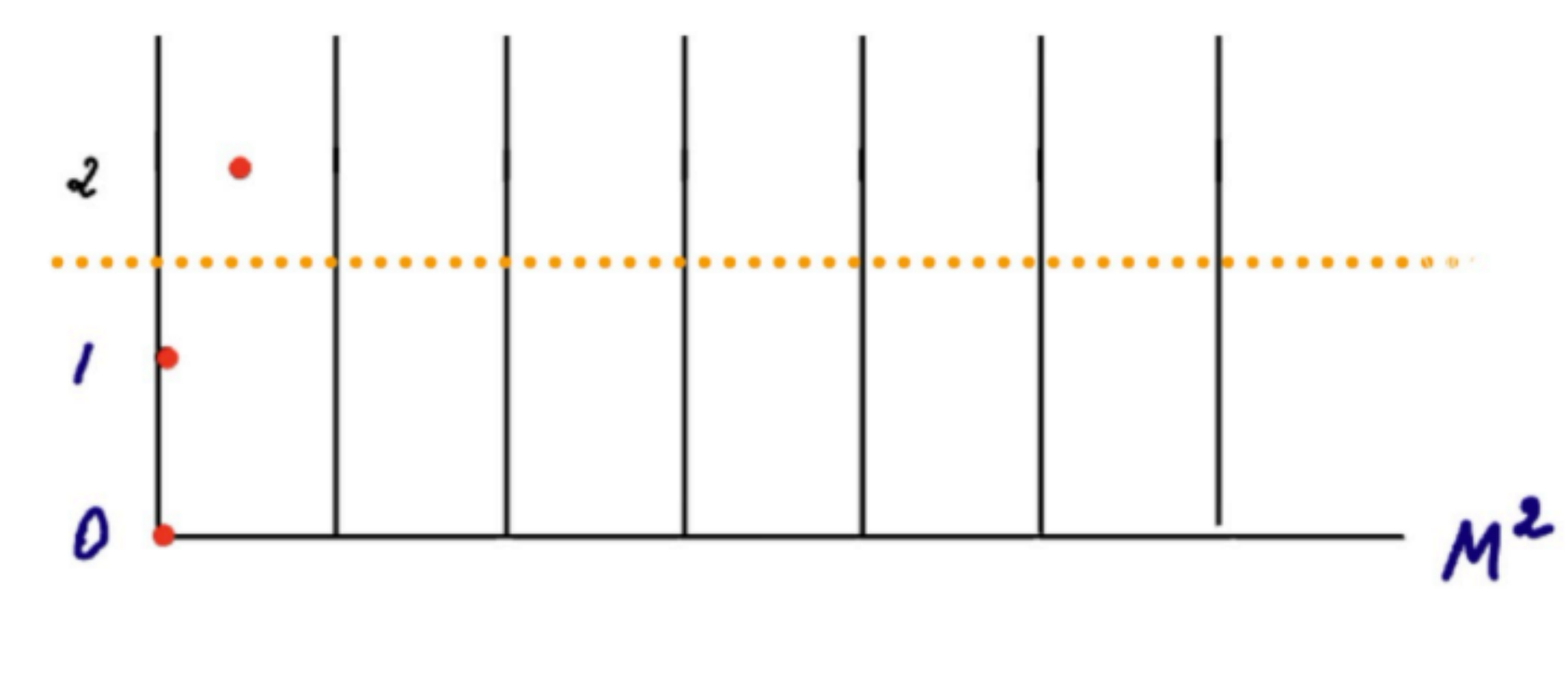}
		\caption{$n=0$ and $n=1$ states.}
		\label{Fig15}
	\end{center}
\end{figure}

The figure \ref{Fig14} shows a sketch of the meson spectrum.
Figure \ref{Fig14} requires some explanation. The integer on the vertical axis differs from the usual integer labeling the 't~Hooft meson spectrum by 2. The meson spectrum in figure \ref{Fig14} begins at $n=2$. This is no more than a redefinition; we have added two states at $n=0$ and $n=1$ (figure \ref{Fig15}). We will call them the photon $(n=0)$ and the graviton $(n=1)$. 
The photon has $CP=-1$ and the graviton has $CP=1$. Photons and gravitons are not dynamically propagating particles in $(1+1)$ dimensions so we have not really added anything to the particle spectrum. But the gauge fields $A_{\mu}$ and $g_{\mu\nu}$ do have meaning. Integrating over $A_{\mu}$ leads to long range $(1+1)$ dimensional Coulomb forces between $U(1)$ electric charges. Indeed such forces are present in the $U(N)$ 't~Hooft model. 

JT-de Sitter is the dimensional reduction of $(2+1)$ dimensional Einstein-Maxwell theory with a positive cosmological constant. The $(2+1)$ dimensional theory does have a propagating photon and electromagnetic forces. When the theory is dimensionally reduced the photon decouples but there is a memory of its existence in electric forces between charges.

It's in this sense that the $U(N)$ 't~Hooft model has a photon.

In a similar way there are forces between heavy states such as Baryons in the 't~Hooft model. These forces may be identified as Gravitational. They are too weak to affect mesons unless the number of mesons is of order $N^{1/2}$, but they affect Baryons. 
\begin{table}[H]
 \begin{center}
   \caption{The 't~Hooft spectrum}
  \begin{tabular}{lll}
    $n=0$ & $CP= -1 $ & \cancel{photon} \\
     ~ & ~ & Long range Coulomb force \\
     ~ & ~ & Confinement of $U(1)$ charge \\
     ~ & ~ & $\checkmark$ feature of the 't~Hooft model \\
     ~ & ~ & ~ \\
    $n=1 $ & $ CP= +1 $  & \cancel{graviton} \\
     ~ & ~ & Long range gravitational force \\
     ~ & ~ & $\checkmark$ feature of the 't~Hooft model \\
     ~ & ~ & ~ \\
    $n \geq 2 $ & $ CP= -,+,-,+, \cdots $ & 't~Hooft-Regge trajectory 
  \end{tabular}
 \end{center}
\end{table}

\subsection{DSSYK Singlet Spectrum for $U(N)$}
Now consider the spectrum of $U(N)$ singlet operators in DSSYK, in the Dirac case.

We don't have a quantitative calculation of the DSSYK spectrum but there are qualitative parallels. There is a basis of singlet operators $M_{n}$ ($n=$ integer),
\be
M_{n} = \psi_{i}^{\dagger} \frac{d^n}{dt^n} \psi_{i} = \psi_{i}^{\dagger} \ddddot{\psi}_{i} ~~~ ({\rm sum ~ on} ~ i) ~.
\ee
As in the 't~Hooft model the CP quantum numbers alternate, 
\be
CP=(-1)^{n+1} ~ .
\ee 

\subsection*{$n=0$}
\be
M_{0} = \psi_{i}^{\dagger} \psi_{i} = Q ~({\rm conserved}) ~ , ~~~~ CP=-1 ~ .
\ee
$M_{0}$ is the total conserved $U(1)$ charge. It has the quantum numbers of the time component of the photon but because it is conserved all its matrix elements are time independent. There is no propagating photon. But photon exchange gives rise to long range confining Coulomb forces. (See \cite{Susskind:2020fiu}.) This parallels the 't~Hooft model.

\subsection*{$n=1$}
\bea
M_{1}  &= & \psi_{i}^{\dagger} \dot{\psi}_{i} \\
~ & = & i \psi_{i}^{\dagger} [H, \psi_{i}] ~ .
\eea
$M_{1}$ can easily be seen to be the Hamiltonian,
\be
M_{1} = \psi_{i}^{\dagger} \dot{\psi}_{i} = H ~~ ({\rm conserved}) ~ .
\ee
Like $Q$, because it is conserved it has no propagating particle. It does have the quantum numbers of the time-time component of the graviton. In particular it is CP even. 

\begin{center}
\begin{minipage}{3cm}
graviton\\
decouples 
\end{minipage}
$\to$ \begin{minipage}{7cm}
Long Range \\
gravitational forces. \\
No propagating mode
\end{minipage}
\end{center}
~ \\
For $n \geq 2$ there is a Regge trajectory of operators with nontrivial dynamics. 

\subsection*{$n=2$} 
\be
\psi_{i}^{\dagger} \ddot{\psi_{i}} = \dot{\psi}_{i}^{\dagger} \dot{\psi}_{i} ~ .
\ee
Massive eta meson, $CP=-1$.

\subsection*{$n=3$} 
\be
\psi_{i}^{\dagger} \dddot{\psi}_{i} ~ .
\ee
$f_{1}$ meson, $CP=+1$. \\
~ \\
\hspace{1cm} {\Huge $\vdots$ }
~\\
~\\
The best test of the \dk-'t~Hooft duality would be to show {\textcolor{red}{\underline{\textcolor{black}{quantitative}}}} agreement between the \dk ~ and the 't~Hooft spectra.

\subsection{Lessons}
Assuming the correctness of all the above what lessons might we draw?
\begin{itemize}
\item Symmetries of the hologram do indeed become bulk gauge symmetries but the symmetries of the hologram can be ``emergent'' only becoming exact in the flat-space limit $N\to\infty$. 

A gauge theory can emerge from a holographic theory with no symmetry at all. 

\item The 't~Hooft model is a string theory,
and therefore so is \dk; a string theory with non propagating photons, gravitons and Regge trajectories of dynamically propagating mesons. But it's not a SUPERSTRING theory. There is no hint of supersymmetry in the 't~Hooft model or in its dual \dk. 

\item Now we come to what may be the most interesting lesson. High energy physics has been laboring under the fine-tuning (of the vacuum energy) problem for well over half a century. It has been the primary motivation for most beyond-the-standard-model theories, from supersymmetric extensions of the standard model to large extra dimensions, none of which solve the problem. A new paradigm is needed. We will argue that the new paradigm is the Holographic Principle.

\end{itemize}

\section{Fine Tuning?}
The argument for fine-tuning is field-theoretic and rests on the apparent existence of quantum corrections to the vacuum energy which depend ultra-sensitively on physics at all scales up to the highest energy, presumably the Planck mass. The argument appears to be very general and not dependent on dimension. 

Why then haven't we encountered it in deriving the SM from \dk? Most interesting: although there is no supersymmetry in either \dk ~ or in the 't~Hooft model there is apparently {\textcolor{red}{\underline{\textcolor{black}{no need for fine-tuning.}}}}

Let's review the usual fine-tuning argument. Corrections to the vacuum energy can be calculated from Feynman diagrams in a fixed background in the bulk theory, i.e., the SM. For example quark loops in the 't~Hooft model.
\begin{figure}[H]
	\begin{center}
		\includegraphics[scale=.4]{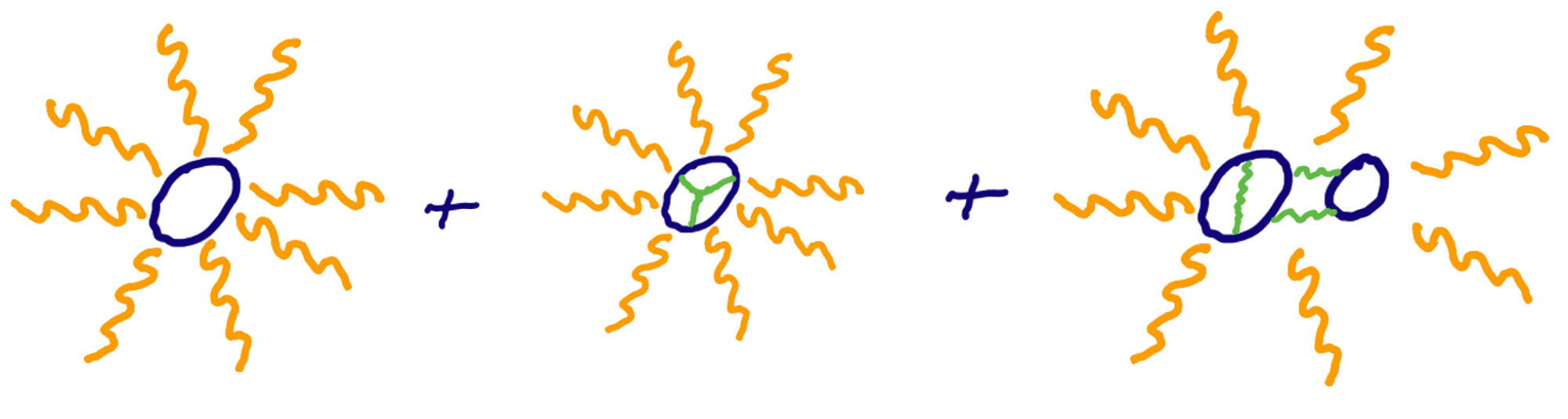}
		\caption{Loop diagrams in the 't~Hooft model.}
		\label{Fig16}
	\end{center}
\end{figure}
Typically the diagrams are divergent and if cut off at the Planck scale give a huge Planckian correction. The diagrams depend on bulk coupling constants which in our case means the gauge-theory coupling $\overline{g}$. Fine-tuning would mean choosing $\overline{g}$ so that the sum of diagrams cancels to one part in a large number.

What is it that is (or is not) fine-tuned? The usual answer is the vacuum energy in Planck units. For example in $D=4$ the dimensionless quantity is 
\be
\epsilon = \rho G^2
\ee
where $\rho$ is the vacuum energy density. But $\epsilon$ is a function of the most fundamental dimensionless measure of de Sitter space: the entropy. In $D=4$ the relationship is 
\be
\epsilon = \frac{3}{8 S} ~ .
\ee
A quantum correction to $\epsilon$ is equivalent to a correction to the de Sitter entropy. In general at fixed temperature interaction can affect the entropy of a system. There are two exceptions. For systems with non-degenerate ground states the entropy at $T_{B} = 0$ is generally zero. The other exception -- the one we are concerned with -- is infinite $T_{B}$. At infinite Boltzmann temperature the entropy is given by the logarithm of the dimension of the Hilbert space independently of the Hamiltonian. Thus for the Majorana and Dirac SYK systems
\bea
S & = & \frac{N}{2} \log 2 ~~~~ {\rm Majorana} \\
S & = & N \log 2  ~~~~~  {\rm Dirac} 
\eea
This is both general and exact. Therefore there can be no corrections to $S$ at infinite $T_{B}$. Since quantum corrections to the vacuum energy are equivalent to corrections to the entropy, corrections to the vacuum energy must vanish assuming the bulk SM derives from a holographic theory at $T_{B}= \infty$. 

\begin{figure}[H]
	\begin{center}
		\includegraphics[scale=.37]{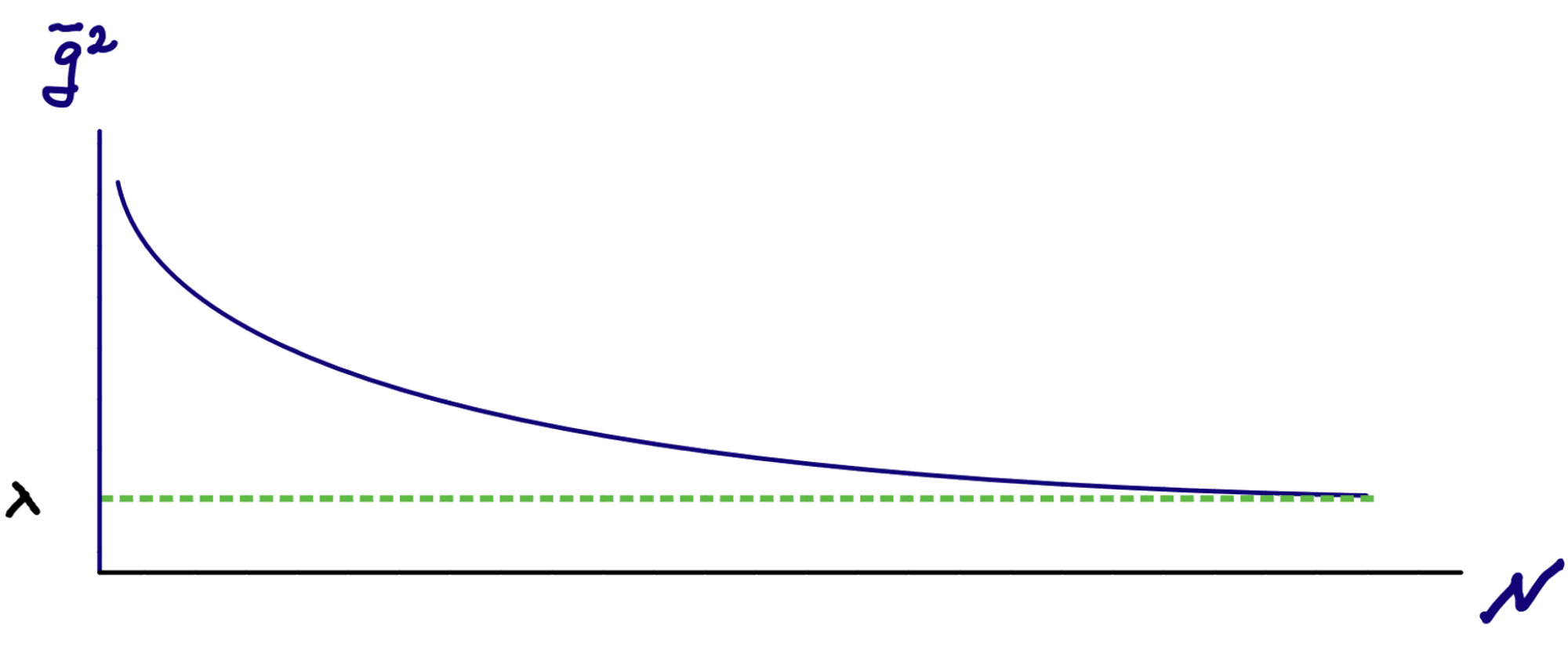}
		\caption{$\overline{g}^2$ as a function of $N$.}
		\label{Fig17}
	\end{center}
\end{figure}

This is a very simple argument but one may wonder what's wrong with the usual argument. To understand what's going on we must take into account that the gauge coupling is a derived quantity in a holographic theory. What we know about it is that in the flat-space limit it is given by 
\be
\overline{g}^2 = \lambda ~.
\ee
However there are corrections to this at finite $N$. Therefore we write 
\be
\overline{g}^2 = \overline{g}^2 (N)
\ee
and 
\be
\overline{g}^2(\infty) = \lambda ~ . 
\ee
A graph of $\overline{g}^2 (N)$ would look something like Fig. \ref{Fig17}. 
~\\
{\textcolor{red}{\underline{\textcolor{black}{Holographic Tuning:}}}} ``Set it ($N$) and forget it.'' \\
Fixing $N$ (and $T_{B}=\infty$) determines the entropy $S$,
\be
S = N \log 2 ~ ,
\ee
and if $N$ is large the vacuum energy will be small. Fixing $N$ also fixes $\overline{g}^2$ at a specific value.

\begin{figure}[H]
	\begin{center}
		\includegraphics[scale=.37]{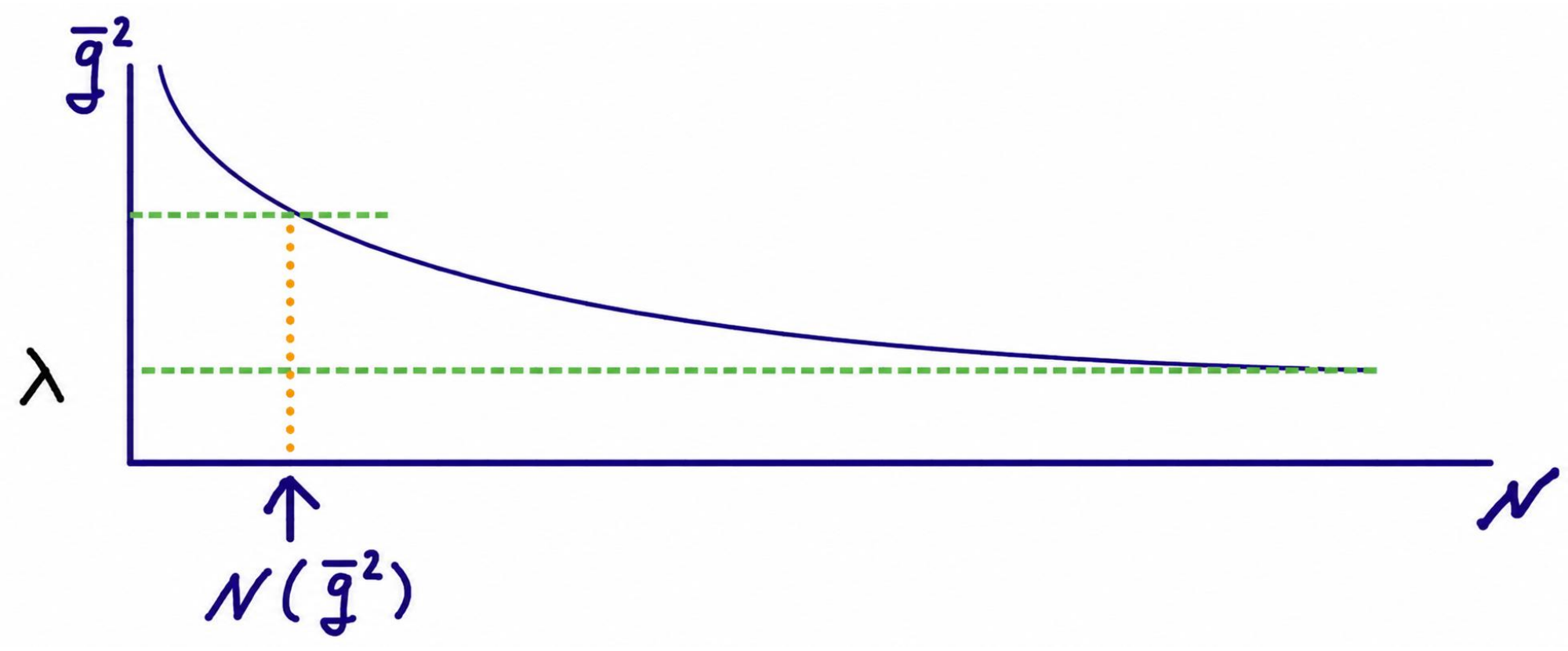}
		\caption{Fix $\overline{g}^2$ and determine the corresponding N.}
		\label{Fig18}
	\end{center}
\end{figure}

By contrast conventional (QFT) tuning fixes $\overline{g}$. One can see from figure \ref{Fig18} that a small shift of $\overline{g}^2$ will require a large shift $N$ and therefore a large change in the vacuum energy.
``Set-it-and-forget-it'' in the holographic theory appears as fine-tuning from the QFT perspective. What is input and what is derived is different in a holographic theory and one based on usual effective QFT. In usual Wilsonian EFT the input is field theoretic parameters in the ultraviolet. In a holographic theory like \dk ~ the input is $N$ and $\lambda$ and are uncorrected at $T_{B} = \infty$. 

Holographic Tuning does not explain why the vacuum energy is so small but it changes the question from: why are the bulk field theory constants fine-tuned? to why is the holographic input $N$ so large?

We've shown you how this works in a very concrete toy model. We don't see any reason why the same principle shouldn't work more generally.

\section{Conclusions} \label{Conclusions}

Over the last few decades a great deal has been learned about quantum gravity, but practically nothing about elementary particles.

The long-standing paradigm for particle theory is Wilsonian QFT. The input is some condensed matter or lattice system at the smallest distance scale. Then through an extraordinary balancing act the input is fine-tuned to give vanishing or almost vanishing ground state energy%
\footnote{This fine-tuning is not the same as the fine-tuning in the search for fixed points of the RG flow. In general the vacuum energy does not vanish at a fixed point.}%
.
	\begin{center}
		\includegraphics[scale=.5]{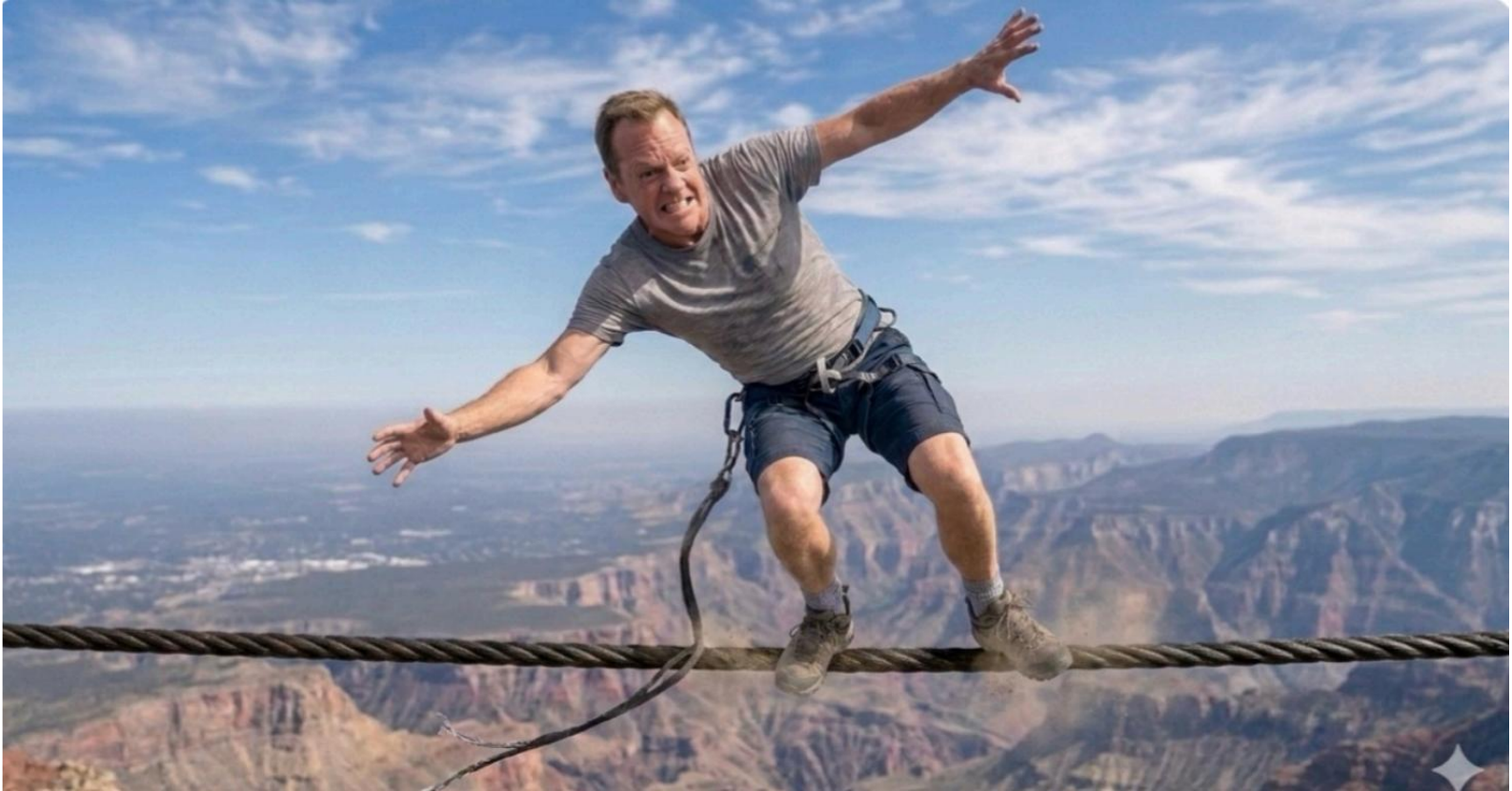}
	\end{center}
One would like to avoid this extreme fine-tuning but the only known mechanism is supersymmetry, a symmetry not present in ``real physics.''

We are proposing a new paradigm for particle physics (not entirely new: see the works of Banks and Fischler which touch on similar issues) based on principles of quantum gravity, specifically the holographic principle. 

The input parameters to the Wilsonian QFT are outputs of the holographic theory. In the case of \dk ~ the input parameters are $N$, $\lambda$, and all importantly $T_{B} = \infty$. By virtue of this last condition the entropy in the flat-space limit is rigorously given (for any Hamiltonian) by, 
\bea
S & = & \frac{N}{2} \log 2 ~~~~ {\rm Majorana} \\
S & = & N \log 2  ~~~~~ {\rm Dirac} 
\eea
$S\to \infty$ implies that the dimensionless vacuum energy goes to zero. There are no corrections: ``Set it and forget it.''
	\begin{center}
		\includegraphics[scale=.5]{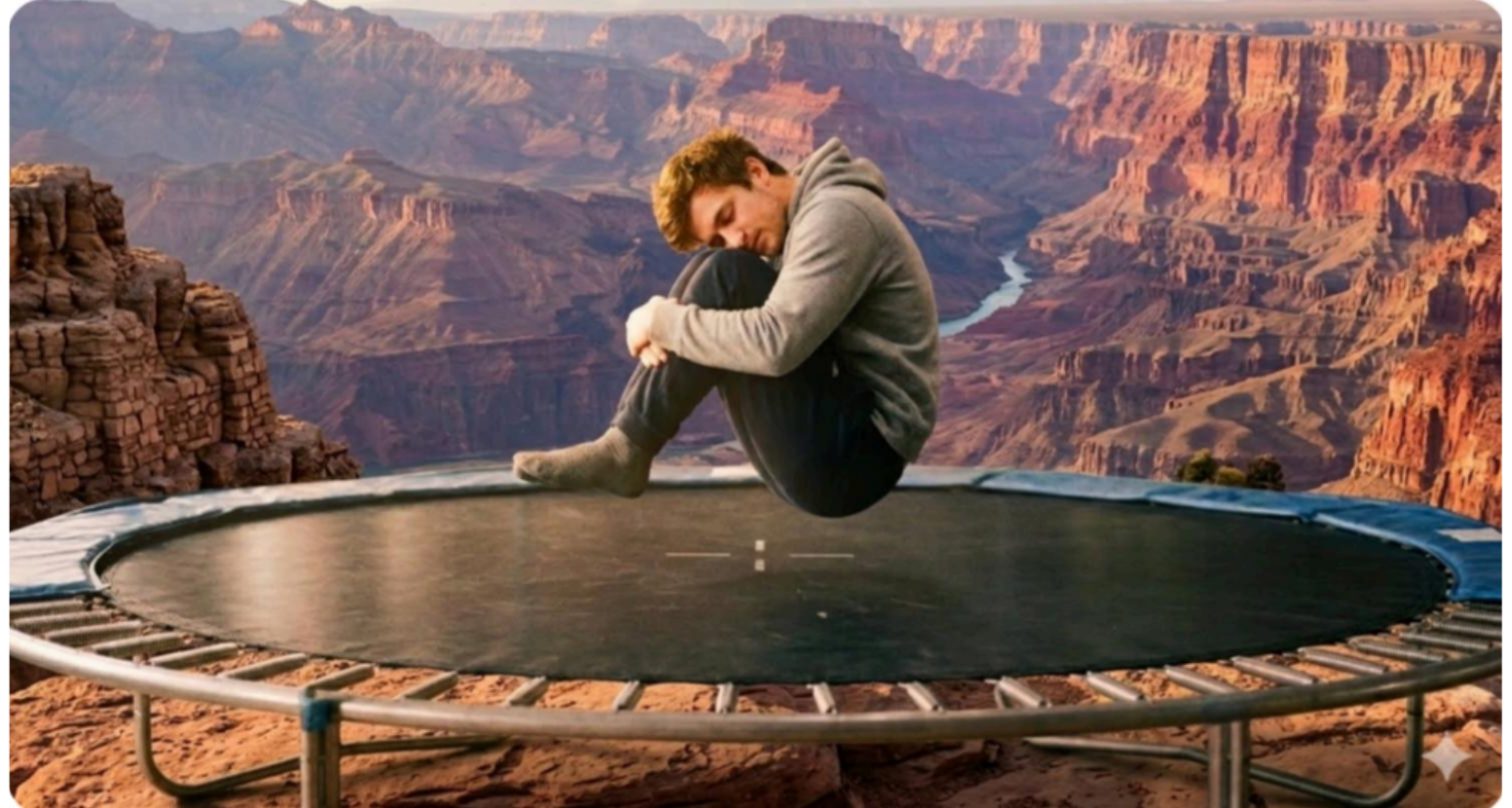}
	\end{center}
Figure \ref{Fig18} illustrates
how the holographic unfine-tuned selection of parameters can appear to be fine-tuned QFT parameter selection. 

At this point we would make no claim about more realistic higher dimensional theories except to say this: \dk ~ appears to contradict the conventional folklore about fine-tuning which was thought to be very general. \dk ~ exposes how holographic theories can evade the folklore. We don't see any reason why the same mechanism cannot work in higher dimensional holographic theories of de Sitter space.

\section*{Acknowledgement}
The work of Y.S. and S.M. is supported in part by MEXT KAKENHI Grant Number 21H05187.

	\end{document}